\documentclass[MSNbibl,nameyear,dvips]{arxstspdf}
\usepackage{flushend}
\usepackage{stfloats}

\volume{27}
\issue{3}
\pubyear{2012}
\firstpage{373}
\lastpage{394}
\doi{10.1214/11-STS378}

\makeatletter

\newcommand{\bA}{\mathbf{A}}
\newcommand{\bH}{\mathbf{H}}
\newcommand{\bI}{\mathbf{I}}
\newcommand{\bM}{\mathbf{M}}
\newcommand{\bQ}{\mathbf{Q}}
\newcommand{\bU}{\mathbf{U}}
\newcommand{\be}{\mathbf{e}}
\newcommand{\bX}{\mathbf{X}}
\newcommand{\cH}{{\cal H}}
\newcommand{\bsigma}{\boldsymbol{\sigma}}
\newcommand{\brho}{\boldsymbol{\rho}}
\newcommand{\cal}{\mathcal}
\makeatother

\begin{document}
\begin{frontmatter}

\title{Quantum Computation and Quantum Information}
\runtitle{Quantum Computation and Quantum Information}

\begin{aug}
\author[a]{\fnms{Yazhen} \snm{Wang}\corref{}\ead[label=e1]{yzwang@stat.wisc.edu}}
\runauthor{Y. Wang}

\affiliation{University of Wisconsin--Madison}

\address[a]{Yazhen Wang is Professor, Department of Statistics,
University of Wisconsin--Madison, Madison, Wisconsin 53706, USA \printead{e1}.}

\end{aug}

%
\begin{abstract}
Quantum computation and quantum information are of great current
interest in computer science, mathematics,
physical sciences and engineering. They will likely lead to a new wave
of technological innovations in communication,
computation and cryptography. As the theory of quantum physics is
fundamentally stochastic,
randomness and uncertainty are deeply rooted in quantum computation,
quantum simulation and quantum information.
Consequently quantum algorithms are random in nature, and quantum
simulation utilizes Monte Carlo techniques extensively.
Thus statistics can play an important role in quantum computation and
quantum simulation, which in turn offer great potential to
revolutionize computational statistics. While only pseudo-random
numbers can be generated by classical computers,
quantum computers are able to produce genuine random numbers; quantum
computers can exponentially or quadratically speed up
median evaluation, Monte Carlo integration and Markov chain simulation.
This paper gives a brief review on quantum computation,
quantum simulation and quantum information.
We introduce the basic concepts of quantum computation and quantum
simulation and present quantum algorithms that are known to be much
faster than the available classic algorithms. We provide a statistical
framework for the analysis of quantum algorithms
and quantum simulation.
\end{abstract}

%
\begin{keyword}
\kwd{Quantum algorithm}
\kwd{quantum bit (qubit)}
\kwd{quantum Fourier transform}
\kwd{quantum information}
\kwd{quantum mechanics}
\kwd{quantum Monte Carlo}
\kwd{quantum probability}
\kwd{quantum simulation}
\kwd{quantum
statistics}.
\end{keyword}

\vspace*{-3pt}
\end{frontmatter}

\section{Introduction}\label{sec1}\vspace*{1pt}

For decades computer hardware has grown in pow\-er approximately
according to
Moore's law, which states that the computer power doubles for constant cost
roughly once every two years. However,
because of the fundamental\vadjust{\goodbreak} difficulties of size in conventional
computer technology,
this dream run is ending.
The conventional approaches to the fabrication of computer technology
are to make
electronic devices smaller and smaller in order to increase the
computer power.
As the sizes of the electronic devices get close to the atomic scale, quantum
effects are starting to interfere in their functioning, and thus the
conventional approaches
run up against the size limit.
One possible way to get around the difficulties is to move to a new
computing paradigm provided by
quantum information science. Quantum information science is based on
the idea of using quantum
devices to perform computation and manipulate and transmit information,
instead of electronic devices following the
laws of classical physics, see \citet{Deu85},\vadjust{\goodbreak} \citet{DiV95}, \citet{Fey81}.
Quantum mechanics and information theory are two of the great scientific
developments and technological revolutions in the 20th century, and
quantum information science is to marry the two previously disparate
fields and form a single unifying viewpoint.
Quantum information science studies the preparation and control of the
quantum states of
physical systems for the purposes of information transmission and
manipulation. It includes quantum computation, quantum communication and
quantum cryptography. This revolutionary field will enable a range of exotic
new devices to be possible. There is now a general
agreement that quantum information science will likely lead to the
creation of a~quantum computer 
to solve problems that could not be efficiently solved on a classical computer.
Already scientists have built rudimentary quantum computers in the
research laboratory to run quantum algorithms
and perform certain calculations. Intensive research efforts
are under way around the world
to investigate a number of technologies that could lead to more
powerful and more prevalent quantum computers in the near future.
It is believed that quantum information and quantum bits are to lead to
a 21st century technological revolution much
as classic information and classic bits did to the 20th century.
Since the theory of quantum mechanics is fundamentally stochastic,
randomness and uncertainty are deeply rooted in
quantum computation and quantum information. As a result, quantum
algorithms are of random nature in the sense that
they yield correct solutions only with some probabilities, and Monte
Carlo methods are widely employed
in quantum simulation.
Thus statistics has an important role to play in quantum computation,
quantum simulation and quantum information.
On the other hand, quantum computation and quantum simulation have
tremendous potential to revolutionize
computational statistics.

A quantum system is generally described by its state, and the state is
mathematically
defined to be a unit vector in some complex Hilbert space. The number
of complex numbers
required to characterize the quantum state
usually grows exponentially with the size of the system, rather than linearly,
as occurs in classical systems. As a consequence,
it takes an exponential number of bits of memory on a classical
computer to store the quantum
state, which puts classical computers
in a difficult position to simulate a quantum system.
On the other hand, nature quantum systems are able to store and keep
track of an exponential\vadjust{\goodbreak}
\mbox{number} of complex numbers and perform data manipulations and
calculations as the systems evolve.
Quantum information science is to grapple with understanding how to
take advantage of the enormous
information hidden in the quantum systems and to harness the immense
potential computational power of
atoms and molecules for the purpose of performing computation and
processing information. Already it
has been shown that quantum algorithms like Grover's search algorithm
and Shor's factoring algorithm
provide great advantage over known classical algorithms.

Contemporary scientific studies often rely on understanding complex quantum
systems, such as those in biochemistry and nanotechnology for the
design of biomolecules and nano-materials.
Quantum simulation is to use computers to simulate a quantum system and
its time evolution.
Classical computers are being used for quantum simulation in designing
novel molecules and creating innovative nano-products.
Quantum computers built upon quantum systems may excel in simulating
naturally occurring
quantum systems, while large quantum systems may be impossible to
simulate in an efficient manner by
classical computers.
A~quantum system with $b$ distinct components may be described with $b$
quantum bits
in a quantum computer, while a classical computer requires $2^b$ bits
of memory to store
its quantum state. This advantage allows quantum computers to
efficiently simulate general
quantum systems that are not efficiently simulatable on classical computers.

In this article we review the concepts of quantum computation and introduce
quantum algorithms and quantum simulation. The quantum algorithms are
known to be much faster than the available
classical algorithms. Statistical analyses of quantum algorithms and
quantum simulation are provided.
We give a brief description on quantum information. 
The article sections start with presentations in broad brushstrokes,
followed by
specific discussions along with some mathematical derivations if
necessary. The intention is to give each topic first an overview and then
a general description and a precise characterization. It is recommended
to focus on the qualitative discussions
but skip the derivations for the readers who would like to get a quick
picture of quantum computation and quantum
simulation.

The rest of the paper proceeds as follows.
Section~\ref{sec2} briefly introduces quantum mechanics, quantum probability and
quantum statistics. Section~\ref{sec3}
reviews basic concepts of quantum computation and entanglement.\vadjust{\goodbreak} Section~\ref{sec4}
illustrates some widely known quantum algorithms and provides
a statistical framework for the study of quantum algorithms. Section~\ref{sec5}
presents quantum simulation and discusses its
statistical analysis. Section~\ref{sec6} gives a short description on quantum
information theory. Section~\ref{sec7} features concluding remarks and lists
some open research problems.

\section{Brief Background Review on Quantum Theory}\label{sec2}
Quantum mechanics has been applied to everything under and inside the
Sun, from chemical reaction 
and superconductor to the structure of DNA and nuclear fusion in stars.
Although the significant difference between classical physics and
quantum physics lies in the quantum prediction of physical
entity when the scale of observations becomes comparable to the atomic
or sub-atomic scale, 
many macroscopic properties of systems can only be fully explained and
understood by quantum physics.
The quantum world is extremely strange, and quantum theory is
completely counterintuitive.
Light waves behave like particles and particles behave like waves (wave
particle duality);
matter can go from one spot to another without moving through the
intermediate space (quantum tunneling);
information can be moved across a vast distance without transmitting it
through the intervening space (quantum teleportation).
Quantum theory provides a
mathematical description of wave particle duality and interaction of
matter and energy. It describes the time evolutions
of physical systems via 
wave functions. The wave functions encapsulate the probabilities that
particles are to be found in a given state at a given time.
For example, the probability of finding a photon in some region is the
square of the modulus
of a wave function, and, since at some point the sum of two wave
functions can be zero but neither wave function is zero,
probabilities appear to cancel out each other in a way totally
unexpected from classical probability.
The intrinsic stochastic nature of quantum theory indicates a deep
connection between quantum mechanics and probability.
Since the main focus of this paper is on quantum computation and
quantum information, we give a~brief description of quantum theory in
this section
to provide some quantum background for the purpose of reviewing quantum
computation and quantum simulation in subsequent sections.
For further reading on the subjects we recommend textbooks by
\citet{SakNap10} at the graduate level and \citet{Gri}\vadjust{\goodbreak}
at the undergraduate level for quantum mechanics, \citet{Hol82}, \citet{Par92} and \citet{Wan94} for quantum probability and quantum stochastic
processes, and \citet{ArtGilGut05} and \citet{BarGilJup03}
for quantum statistics.

\subsection{Hilbert Space and Operator}\label{sec2.1}

For the sake of simplicity we choose to work with comparatively easy
finite-dimensional situations.
Denote by $\mathbb{C}$ the set of all complex numbers.
We start with vector space in linear algebra. A simple example of
vector space is $\mathbb{C}^k$
consisting of all $k$-tuples of complex numbers $(z_1,\ldots,z_k)$.
The elements of a vector space are
called vectors. As in quantum mechanics and quantum computation, we use
Dirac notations
$|\cdot\rangle$ (which is called ket) and $\langle\cdot|$ (which is
called bra) to indicate that
the objects are column vectors or row vectors in the vector space, respectively.
Denote by superscripts $*$, $\prime$ and $\dagger$ the conjugate of a
complex number, the transpose
of a vector or matrix, and conjugate transpose operation, respectively.
We define an inner product on the
vector space to be a function that takes as input two vectors from the
vector space and produces a
complex number as output. For $|u \rangle$ and $|v \rangle$ in the
vector space, we denote their inner product
by $\langle u |v \rangle$. The inner product must satisfy
(i) conjugate symmetry, $\langle u |v \rangle= (\langle v | u \rangle
)^*$; (ii) linearity in
the second argument, $\langle u |v + w \rangle= \langle u |v \rangle+
\langle u |w \rangle$;
(iii) positive-definiteness, $\langle u |u \rangle\geq0$ with
equality only for $u = 0$.
For example, $\mathbb{C}^k$ has a natural inner product
\[
\langle u | v \rangle= \sum_{j=1}^k u_j^* v_j =(u_1^*,\ldots, u^*_k)
(v_1,\ldots, v_k)^\prime,
\]
where $\langle u| = (u_1, \ldots, u_k)$ and $|v \rangle= (v_1, \ldots
, v_k)^\prime$. An inner product
induces a norm
$ \| u \| = \sqrt{ \langle u | u \rangle}, $ and a~distance $\| u -
v\|$ between $| u \rangle$ and $| v \rangle$.
For the finite-dimensional case, a Hilbert space $\cH$ is simply a
vector space with an inner product.

An operator $\bA$ on $\cH$, denoted by $\bA(|u\rangle)$ for \mbox{$|u
\rangle\in\cH$}, is a function mapping from
$\cH$ to $\cH$ that satisfies
$\bA(a |u \rangle+ b |v \rangle) = a \bA(|u \rangle) + b \bA(|v
\rangle)$ for
any \mbox{$|u \rangle, |v \rangle\in\cH$} and $a, b \in\mathbb{C}$.
We can represent an operator through a matrix.
Suppose that $\bA$ is an operator on $\cH$ and $\be_1, \ldots, \be
_k$ form an orthonormal
basis in $\cH$. Then each $\bA(|\be_j\rangle) \in\cH$ and there
exists a unique $k \times k$ matrix $(a_{j\ell})$
such that $ \bA(|\be_j\rangle)=\sum_{\ell=1}^k |\be_\ell\rangle
a_{\ell j} $, $j=1,\ldots, k$. We will identify operator $\bA$ with
matrix $(a_{j\ell})$ and use $\bA$ for both operator and matrix
$(a_{j\ell})$. An operator $\bA$ on $\cH$ is said to be self-adjoint
if its corresponding\vadjust{\goodbreak} matrix $\bA$ is Hermitian, that is, $\bA= \bA
^\dagger$.
We also refer to self-adjoint operators as Hermitian operators.
An operator $\bU$ is said to be unitary if its corresponding matrix
$\bU$ is unitary, that is,
$\bU\bU^\dagger= \bU^\dagger\bU= \bI$. We say an operator $\bA$
is semi-positive (or positive) definite
if its corresponding matrix $\bA$ is semi-positive (or positive)
definite, that is, $\langle u| \brho|u \rangle\geq0$
for $|u \rangle\in{\cal H}$ (or $\langle u| \brho|u \rangle\geq0$
for $|u \rangle\in{\cal H}$
with equality only for $|u \rangle=0$). The trace of an operator $\bA
$, denoted by $\operatorname{Tr}(\bA)$, is defined
to be the trace of its corresponding matrix $\bA=(a_{j\ell})$, that
is, $\operatorname{Tr}(\bA)=\sum_{j=1}^k a_{jj}$.

\subsection{Quantum System}\label{sec2.2}
Quantum mechanics depicts phenomena at microscopic level such as position
and momentum of an individual particle like an atom or electron, spin
of an
electron, detection of light photons, and the emission and absorption of
light by atoms. Unlike classical mechanics where physical entities
like position and momentum can be measured 
precisely, the theory of quantum mechanics is intrinsically stochastic
in a sense
that we can only make probabilistic prediction about the results of the
measurements performed.\looseness=-1 

Quantum mechanics is mathematically described by a Hilbert space $\cal H$
and self-adjoint operators on~$\cal H$.
A~quantum system is completely characterized by its state and the time
evolution of the state.
A~state is defined to be a unit vector in $\cH$. Let $|\psi(t)\rangle
$ be the state of the quantum
system at time~$t$, which is also referred to as a wave function.
The states $|\psi(t_1)\rangle$ and $|\psi(t_2)\rangle$ at $t_1$ and
$t_2$ are connected through
$ |\psi(t_2)\rangle= \bU(t_1, t_2) |\psi(t_1)\rangle, $ where
$\bU(t_1,t_2)$ is a unitary operator
depending only on time $t_1$ and $t_2$.
In fact, there exists a self-adjoint operator $\bH$, which is known as
the Hamiltonian of the
quantum system, such that
$ \bU(t_1,t_2) = \exp[ -i  \bH (t_2-t_1)]. $
With Hamiltonian $\bH$, we may describe
the continuous time evolution of $|\psi(t)\rangle$ by the Schr\"odinger equation
%
\begin{equation} \label{schrodinger1}
i \frac{\partial|\psi(t) \rangle}{\partial t} = \bH|\psi(t)
\rangle, \quad i = \sqrt{-1}.
\end{equation}

Alternatively a quantum system can be described by a density operator
(or density matrix).
A density operator $\brho$ is an operator on $\cH$ which (1) is self-adjoint;
(2) is semi-positive definite; (3) has unit trace [i.e., $\operatorname
{Tr}(\brho)=1$].
Following the convention in quantum information science, we reserve
notation $\brho$ for state, density operator or density matrix.
A state is often classified as a pure state or an ensemble of pure
states. A pure state is a unit vector $|\psi\rangle$
in $\cal H$, which corresponds to a density operator $\brho= |\psi
\rangle\langle\psi|$, and
an ensemble of\vadjust{\goodbreak} pure states corresponds to the case that the quantum
system is in one of
states $|\psi_j \rangle$, $j=1,\ldots, J$, with probability $p_j$
being in state
$|\psi_j \rangle$, and the corresponding density operator
%
\begin{equation} \label{density}
\brho= \sum_{j=1}^J p_j  |\psi_j \rangle\langle\psi_j |.
\end{equation}
See \citet{Gri}, \citet{SakNap10} and \citet{Sha94}.

\subsection{Quantum Probability}\label{sec2.3}
We can test the theory of quantum mechanics by checking its predictions
with experiments of
performing measurements on quantum systems in the laboratory.
The usual quantum measurements are on observables such as
position, momentum, spin, and so on, where an \textit{observable} $\bX$
is defined
as a~self-adjoint operator on Hilbert space $\cal H$. The \textit{observable} definition
is motivated from the fact that the eigenvalues of self-adjoint
operators are real.
Assume that an observable $\bX$ has a discrete spectrum with the
following diagonal form
%
\begin{equation} \label{diagonal}
\bX= \sum_{a=1}^p x_a  \bQ_a,
\end{equation}
where $x_a$ are real eigenvalues of $\bX$ and $\bQ_a$ are the corresponding
one-dimensional projections onto the orthogonal eigenvectors of $\bX$.
Consider such an observable in the quantum system prepared in state~$\brho$.
\mbox{Measure} space $(\Omega, {\cal F})$ is used to describe possible
measurement outcomes of the observable,
and the result of the measurement is a random variable on $(\Omega,
{\cal F})$ with
probability distribution $P_{\brho}$.
We denote by~$X$ the result of the measurement of observable~$\bX$
given by (\ref{diagonal}).
Then~$X$ is a random variable taking values in $\{x_1, x_2, \ldots, \}$,
and under pure state $|\psi\rangle$, the probability that measurement
outcome $x_a$ occurs is defined to be
\begin{eqnarray*}
P(a) &=& P_{\brho}(X = x_a)\\\
&=& \langle\psi|\bQ_a | \psi\rangle
=\operatorname{Tr}(\bQ_a  |\psi\rangle \langle\psi| ),   \quad
a=1, 2, \ldots.
\end{eqnarray*}
With the probability we derive the expectation under pure state $| \psi
\rangle$,
\begin{eqnarray*}
\mathbb{E}_\psi(\bX) &=& \sum_a x_a P(a) = \sum_a x_a   \langle
\psi| \bQ_a |\psi\rangle\\
&=&
\langle\psi| \bX|\psi\rangle=\operatorname{Tr}(\bX |\psi
\rangle \langle\psi| ).
\end{eqnarray*}
Note the 
difference between an observable $\bX$ which is a Hermitian matrix
and its measurement result $X$ which is a real-valued random variable.

Measuring observable $\bX$ will alter the state of the quantum
system\vadjust{\goodbreak}
(\cite*{Kie04}; \cite*{von55}).
If the quantum system is prepared with initial state $|\psi\rangle$,
the state of the system after the measurement result $x_a$ is defined
to be
%
\begin{equation} \label{post-state}
\frac{\bQ_a |\psi\rangle}{\sqrt{P(a)} }.
\end{equation}

For an ensemble state with density operator $\brho$ given by (\ref{density}),
if the quantum state is $|\psi_j \rangle$, the probability that
result $x_a$ occurs is
\[
P(a|j) = \langle\psi_j|\bQ_a|\psi_j\rangle=
\operatorname{Tr}(\bQ_a  |\psi_j\rangle \langle\psi_j| ).
\]
Applying the law of total probability, we obtain that under state
$\brho$,
the probability that $x_a$ occurs is equal to
\begin{eqnarray*}
P(a) &=& P_{\brho}(X=x_a) = \sum_{j=1}^J p_j P(a|j)\\
 &=& \sum_{j=1}^J
p_j \operatorname{Tr}(\bQ_a  |\psi_j\rangle \langle\psi_j| ) =
\operatorname{Tr}(\bQ_a  \brho).
\end{eqnarray*}
The expectation of $\bX$ under state $\brho$,
\begin{eqnarray*}
\mathbb{E}_{\brho} [\bX] &=& \sum_{a=1}^p x_a  P_{\brho}[ X = x_a]
= \sum_{a=1}^p x_a  \operatorname{Tr}(\bQ_a  \brho)\\
&=& \operatorname{tr}( \bX \brho), 
\end{eqnarray*}
and variance
\[
\operatorname{Var}_{\brho} [\bX] = \operatorname{tr}[\bX^2 \brho
] - (\operatorname{tr}[\bX \brho])^2.
\]

We may derive the density operator of the quantum system after
obtaining the measurement result~$x_a$ by conditional probability
arguments as follows.
If the quantum system is in pure state $|\psi_j\rangle$ before the
measurement, the quantum state after measurement result $x_a$ has
occurred is
\[
|\psi_j^a \rangle= \frac{\bQ_a |\psi_j \rangle}{\sqrt{P(a|j)} }.
\]
If the quantum state is $\brho$ before the measurement, after
observing measurement outcome $x_a$ we have the following ensemble of
states: the quantum system is in pure state
$|\psi_j^a \rangle$ with probability $P(j|a)$, where Bayes's theorem shows
\[
P(j|a) 
= p_j P(a|j) /P(a).
\]
Thus after measurement $x_a$ the density operator for the ensemble
state is given by
\begin{eqnarray*}
\brho_a& =& \sum_{j=1}^J P(j|a)  |\psi_j^a\rangle\langle\psi_j^a|\\
&=& \sum_{j=1}^J P(j|a)  \frac{\bQ_a |\psi_j \rangle\langle\psi_j|
\bQ_a
}{P(a|j) }\\
&=& \sum_{j=1}^J p_j  \frac{\bQ_a |\psi_j \rangle\langle\psi_j|
\bQ_a
}{P(a) } = \frac{\bQ_a  \brho\bQ_a}{\operatorname{Tr}(\bQ_a
\brho)}.
\end{eqnarray*}
See \citet{Hol82}, \citet{Par92} and \citet{SakNap10}.

\subsection{Quantum Statistics}\label{sec2.4}
For a given quantum system, it is very important but difficult to know
its state. If we do not know in
advance the state of the quantum system, we may infer the quantum state
by the measurement results of
some observables obtained from the quantum system and show that a
certain state has been created.
In statistical terminology, we want to estimate density matrix $\brho$ based
on measurements on an often large number of systems which are
identically prepared in
the state $\brho$. That is, after measuring observables on some
identical quantum systems, we can
make statistical inference about probability distribution $P_{\brho}$
of the measurements and
thus indirectly about density matrix $\brho$.
In the literature of quantum physics, quantum tomography is referred to
as the reconstruction of
the underlying density matrix $\brho$ by probing identically prepared
quantum systems from some different angles.
Specifically, suppose that we perform measurements of observables on
identically prepared quantum systems
in an unknown state $\brho$ and obtain measurement results $X_1,
\ldots, X_n$.
Assume that $\brho$ is known up to some unknown parameter $\theta$;
then $X_1, \ldots, X_n$ are i.i.d. observations
with distributions $P_{\brho}$ which depend on $\theta$. This gives a
quantum parametric
statistical model. We may then define quantum likelihood and Fisher
quantum information and establish
quantum point estimation and quantum hypothesis testing theory.
Alternatively we may model~$\brho$ nonparametrically by assuming
that~$\brho$ is an infinite matrix
and then use nonparametric methods to estimate the density matrix.
For details see \citet{ArtGilGut05}, \citet{BarGilJup03}, \citet{ButGutArt07} and \citet{NusSzk09}.

\section{Quantum Computing Concepts}\label{sec3}

Unlike classical computers using transistors to\break  crunch the ones and
zeroes individually,
quantum computers can handle both one and zero simultaneously via what
are known as superposition\vadjust{\goodbreak}
quantum states. A superposition state is a state of matter which we may
think of as both one and
zero at the same time. Quantum computers use the strange superposition
states and quantum entanglements
to do the trick of performing simultaneous calculations
and extracting the calculated results. The spooky phenomena of quantum
entanglement and superposition
are the key that enables quantum computers to be superfast and vastly
outperform classical computers.

\subsection{Quantum Bit}\label{sec3.1}
Analogous to the fundamental concept of bit in classical computation
and classical information, we have its
counterpart, quantum bit, in quantum computation and quantum
information. Quantum bit is called qubit
for short. Just like a classical bit with state either $0$ or $1$, a
qubit has states $|0\rangle$
and $|1\rangle$. However, the real difference between a bit and
a~qubit is that
besides states $|0\rangle$ and $|1\rangle$, a qubit may take the
superposition states,
\[
|\psi\rangle= \alpha_0  |0\rangle+ \alpha_1  |1\rangle,
\]
where $\alpha_0$ and $\alpha_1$ are complex numbers and called
amplitudes satisfying
$|\alpha_0|^2 + |\alpha_1|^2=1$.
That is, the states of a qubit are unit vectors in a two-dimensional
complex vector space,
and states $|0\rangle$ and $|1\rangle$ consist of an orthonormal
basis for the space and are
often referred to as computational basis states. For a classical bit we
can examine it
to determine whether it is in the state $0$ or $1$. However, for a
qubit we cannot determine its state
and find the values of $\alpha_0$ and $\alpha_1$ by examining it. The
stochastic nature of quantum theory
shows that we can measure a qubit and obtain either\vspace*{1pt} the result $0$,
with probability $|\alpha_0|^2$, or
the result $1$, with probability $|\alpha_1|^2$. Physical experiments
have realized qubits as physical objects
in different physical systems, such as the two states of an electron
orbiting a single atom, the two
different polarizations of a photon, or the alignment of a nuclear spin
in a uniform magnetic field.
Consider the case of atom model by corresponding
$|0\rangle$ and $|1\rangle$ with the so-called ``ground'' and
``excited'' states of the electron,
respectively. As the atom is shined by light with suitable energy and
for a proper amount of time,
the electron can be moved from the $|0\rangle$ state to the $|1\rangle
$ state and vice versa. Furthermore,
by shortening the length of time shining the light on the atom, we may
move the electron initially
in the state $|0\rangle$ to ``halfway'' between $|0\rangle$ and
$|1\rangle$, say, into a state
$(|0\rangle+ |1\rangle)/\sqrt{2}$.

Note that for qubit state $|\psi\rangle$ the only measurable
quantities are the probabilities $|\alpha_0|^2$ and
$|\alpha_1|^2$;\vadjust{\goodbreak}
since $|e^{i \theta} \alpha_x|^2=|\alpha_x|^2$, where $x=0,1$,
$i=\sqrt{-1}$, and~$\theta$ is a~real number, from the viewpoint of the
qubit measurements, states $e^{i \theta} |\psi\rangle$ and
$|\psi\rangle$ are identical. That is, multiplying a qubit state by a
global phase factor~$e^{i \theta}$
bears no observational consequence.

Note the distinction between superposition states and probability
mixtures (or ensemble of pure states defined in Section~\ref{sec2.1}).
Consider superposition $(|0\rangle+ |1\rangle)/\sqrt{2}$ as a pure
state. Its density matrix is given by
\begin{eqnarray*}
&&\tfrac{1}{2} (|0\rangle+ |1\rangle)(\langle0| + \langle1|)\\
&& \quad  =
\tfrac{1}{2} (|0\rangle\langle0|+ |1\rangle\langle1|) + \tfrac
{1}{2} (|0\rangle\langle1|+ |1\rangle\langle0|),
\end{eqnarray*}
while the first term on the right-hand side of the above equation
corresponds to the ensemble of pure states $|0\rangle$ and
$|1\rangle$, that is, a probabilistic mixture of states $|0\rangle$
and $|1\rangle$ with equal probability.

Similar to classic bits, we can define multiple  qu\-bits. The states of
$b$ qubits are unit vectors in
a $2^b$-dimensional complex vector space with $2^b$ computational basis
states of the form
$|x_1 x_2\cdots x_b \rangle$, $x_j=0$ or~$1$, $j=1,\ldots, b$.
For example, the states of two qubits are unit vectors
in a four-dimensional complex vector space, with four computational
basis\break states labeled by
$|00\rangle$, $|01\rangle$, $|10\rangle$ and $|11\rangle$. The
computational basis states
$|00\rangle$, $|01\rangle$, $|10\rangle$ and $|11\rangle$ generate
the four-dimensional complex
vector space, and the superposi\-tion states are all unit vector in the
space with the forms
\[
|\psi\rangle= \alpha_{00} |00\rangle+ \alpha_{01} |01\rangle+
\alpha_{10}  |10\rangle+
\alpha_{11}  |11\rangle,
\]
where amplitudes $\alpha_x$ are complex numbers satisfying
$|\alpha_{00}|^2 + |\alpha_{01}|^2 + |\alpha_{10}|^2 + |\alpha
_{11}|^2 =1$. As in the
single qubit case, when two qubits are measured we get result $x$ being
one of $00, 01, 10, 11$,
with probability $|\alpha_x|^2$. 
Moreover, we may measure just the first qubit of the two-qubit system
and obtain either the result $0$,
with probability $|\alpha_{00}|^2 +|\alpha_{01}|^2$, or the result
$1$, with probability $|\alpha_{10}|^2 +|\alpha_{11}|^2$.
As quantum measuring changes the quantum state, if the measurement
result on the first qubit is $0$, after the measurement
the qubits are in the state
%
\begin{equation} \label{post-qubit}
\frac{ \alpha_{00} |00\rangle+ \alpha_{01} |01\rangle}{\sqrt
{|\alpha_{00}|^2 + |\alpha_{01}|^2}}.
\end{equation}

A qubit is the simplest quantum system. The quantum system of $b$
qubits is described by a $2^b$-dimension\-al complex vector space
with each superposition state specified by $2^b$ amplitudes. As $2^b$
increases exponentially in $b$, it is very easy for such a
system to have an enormously big vector space.
A quantum system with\vadjust{\goodbreak} even a few dozens of ``qubits'' will strain the
resources of even the largest supercomputers.
Consider a quantum system of $50$ qubits. $2^{50} \approx10^{15}$
complex amplitudes are needed to specify its quantum states.
With $128$ bits of precision, it requires approximately $32$ thousand
terabytes of information to store all $10^{15}$
complex amplitudes. Such storage capacity may be available in future
supercomputers.
For a quantum system with $b=500$ qubits we need to specify $2^{500}$
complex amplitudes for its states. It is
unimaginable to store all $2^{500}$ complex numbers in any classical
computers. In principle, a quantum system
with only a few hundred atoms can manage such an enormous amount of
data and execute calculations as the system
evolves. Quantum computation and quantum information are to find ways
to utilize the immense potential
computational power in quantum systems.

\subsection{Quantum Circuit Model}\label{sec3.2}
As a classical computer is built from an electrical circuit consisting of
wires for
carrying information around the circuit and logic gates for performing
simple computational tasks,
a quantum computer can be created from a quantum circuit with quantum
gates to
perform quantum computation and manipulate quantum information.
A number of physical systems are being investigated for building
quantum computers. These include
optical photon, optical cavity quantum electrodynamics, ion traps,
nuclear magnetic resonance with molecules,
quantum dots, and superconductors (\cite*{NieChu00}). In fact,
primitive solid-state quantum processors have been created in research
laboratories to run quantum algorithms (\cite*{DiCetal09};
\cite*{Johetal};
\cite*{Maretal};
\cite*{Sayetal11}).
The circuit model is particularly important in quantum computation and
quantum information,
and a quantum computer is often synonymous with the quantum circuit model.
A quantum circuit operates on $b$ qubits for some integer $b$. The
state takes a form of
$|x_1\cdots x_b\rangle$, with state space being a $2^b$-dimensional
complex Hilbert space.
When $x_i=0$ or $1$, states $|x_1\cdots x_b\rangle$ are the
computational basis states of the quantum computer
and often written as $|x\rangle$, where $x$ is the integer with binary
representation $x_1\cdots x_b$.

As a classical logic
gate converts classical bits from one form to another such as
$0\rightarrow1$
and $1 \rightarrow0$, a~quantum gate operates on qubits. Quantum
mechanics dictates that quantum gates
operating on $b$ qubits are $2^b$ by $2^b$ unitary matrices on the
$2^b$-dimensional Hilbert space.
For example, a Hadamard gate on one\vadjust{\goodbreak} qubit is the $2 \times2$
unitary
matrix that realizes the following transformation:
\[
|0\rangle\rightarrow\frac{|0 \rangle+ |1 \rangle}{\sqrt{2}},
\quad
|1\rangle\rightarrow\frac{|0 \rangle- |1 \rangle}{\sqrt{2}}.
\]
Consider another important gate on two qubits which is called
control-NOT gate. It takes the two input qubits as
control qubit and target qubit, respectively, and the output target
qubit of the gate retains the input target qubit if the control qubit
is $|0\rangle$ and is flipped if the control qubit is $|1\rangle$,
that is,
\begin{eqnarray*}
|00\rangle&\rightarrow&|00\rangle, \quad     |01\rangle\rightarrow
|01\rangle, \\
|10\rangle&\rightarrow&|11\rangle,  \quad  |11 \rangle\rightarrow
|10\rangle.
\end{eqnarray*}
Generally for any single qubit unitary operation $\bU$, a control-$\bU
$ gate is a two-qubit gate, with
one control qubit and one target qubit. If the control qubit is
$|1\rangle$, $\bU$ is applied to the
target qubit; if the control qubit is $|0\rangle$, the target qubit is
left alone, that is,
\begin{eqnarray*}
|0\rangle|0\rangle&\rightarrow&|0\rangle|0\rangle,  \quad  |0\rangle
|1\rangle\rightarrow|0\rangle|1\rangle,  \\
|1\rangle|0\rangle&\rightarrow&|1\rangle\bU|0\rangle,  \quad
|1\rangle|1 \rangle\rightarrow|1\rangle\bU|1\rangle.
\end{eqnarray*}
If $f(x)$ maps $\{0,1\}^b$ onto $\{0,1\}$, we define a unitary
transformation $\bU_f$ that
operates on $b+1$ qubit state
%
\begin{equation} \label{Uf}
|x, y \rangle\rightarrow|x, y\oplus f(x)\rangle,
\end{equation}
where $x=x_1\cdots x_b$ with $x_j=0$ or $1$ is the data register, $y=0$
or $1$ is the
target register, $\oplus$ denotes additional modulo $2$.
If $y=0$, after the transformation $\bU_f$, the state of the last
qubit is the value of~$f(x)$.

\subsection{Entanglement}\label{sec3.3}
Quantum entanglement is one of the most mind-bending creatures known to
science. It is referred to as the phenomenon that
two qubits behave like twins that are connected by an invisible wave to
share each other's properties.



\subsubsection{Bell states}\label{sec3.3.1}
Consider a quantum gate on two-qubit basis states $|00 \rangle$, $|01
\rangle$, $|10 \rangle$ and $|11 \rangle$ that
is composed of a Hadamard gate on the first qubit and then is followed
by a control-NOT gate. The output states of the gate are as follows:
\begin{eqnarray*}
|00\rangle&\rightarrow&\frac{|00\rangle+ |11\rangle}{\sqrt{2}},  \quad
|01\rangle\rightarrow\frac{|01\rangle+ |10\rangle}{\sqrt{2}},
\\
|10\rangle&\rightarrow&\frac{|00\rangle- |11\rangle}{\sqrt{2}},  \quad
|11\rangle\rightarrow\frac{|01\rangle- |10\rangle}{\sqrt{2}}.
\end{eqnarray*}
Physicists Bell, Einstein, Podolsky and Rosen discovered the amazing
properties of these four states, which are often referred
to as the Bell states, EPR states or EPR pairs (\cite*{Bel64};
\cite*{EinPodRos35}).
In\vadjust{\goodbreak} general states such as these four states that cannot be expressed as
products of some single qubits are called entangled states.
Entangled states, which are not fully understood in quantum physics,
have remarkable properties.

For the two-qubit system consider a Bell state 
\[
|\psi\rangle= \frac{|01\rangle- |10\rangle}{\sqrt{2}},
\]
and an observable
\[
\bM= a_x \bsigma_x + a_y \bsigma_y + a_z \bsigma_z,
\]
where $(a_x, a_y, a_z)$ is a real unit vector (i.e.,
$a_z^2+a_y^2+a_z^2=1$), and $\bsigma_x$, $\bsigma_y$ and $\bsigma_z$
are Pauli matrices given by
%
\begin{eqnarray} \label{Pauli}
\bsigma_x &=& \left(
\matrix{ 0 & 1 \cr 1 & 0
}
\right), \quad
\bsigma_y = \left(
\matrix{ 0 & -i \cr i & 0
}
\right),  \nonumber
\\[-8pt]
\\[-8pt]
\bsigma_z &=& \left(
\matrix{ 1 & 0 \cr 0 & -1
}
\right),  \quad  i = \sqrt{-1}.
\nonumber
\end{eqnarray}
It is easy to show that $\bM$ has eigenvalues $\pm1$ for any real
unit vector $(a_x, a_y, a_z)$.
If measuring observable $\bM$ on each qubit of $|\psi\rangle$, we
will obtain a~measurement result of
$+1$ or $-1$. Surprisingly, no matter what choice of $(a_x, a_y, a_z)$,
the measurement results on the two qubits
are always opposite of each other, that is, when the first qubit
measurement is $-1$, then the second qubit
measurement will be $+1$, and vice versa.

The two-qubit system can be realized by the spins of two particles, and
the measurement of $\bM$ is referred to as
a measurement of spin along the $(a_x, a_y,\break a_z)$ axis. After the
two-particle system is prepared in the Bell
state $|\psi\rangle$, the two particles drift far apart. Alice
measures the spin of the first particle and Bob measures
the spin of the second particle. The above opposite measurement
phenomenon corresponds to that due to the entangled state $|\psi
\rangle$,
if Alice gets a~result $+1$ from her spin measurement on the first
particle, then the state of the
system immediately jumps to the untangled state so that the second
particle now has definite spin state and Bob's spin measurement
on the second particle gives definite result $-1$. This phenomenon is
often referred to as anti-correlation in entanglement experiments
(\cite*{Neuetal08};
\cite*{SakNap10}).

The mathematical arguments for the anti-correla\-tion phenomenon are as
follows. The measurement of $\bM$ on the first (or second) qubit of
$|\psi\rangle$
corresponds to the spin measurement of Alice's (or Bob's) particle
along the $(a_x, a_y, a_z)$ axis in the above two-particle spin model.
From Sections~\ref{sec2.3} and~\ref{sec3.1} we have that
the two-qubit system is described by the Bell state $|\psi\rangle$ in
$\mathbb{C}^4$; 
measuring $\bM$ on the first qubit of $|\psi\rangle$ means
performing measurement on observable $\bM\otimes\bI$ in the Bell
state $|\psi\rangle$,
which alters the quantum state of the two-qubit system; measuring $\bM
$ on the second qubit corresponds to measuring observable $\bI\otimes
\bM$ in
the altered quantum state, where $\bI$ is the 2 by 2 identity matrix,
and $\bM\otimes\bI$ and $\bI\otimes\bM$ are matrix tensor products.

Denote by
$|\varphi_{\pm} \rangle$ the two orthonormal eigenvectors of $\bM$
corresponding to eigenvalues $\pm1$, respectively,
and let $\bQ_{\pm}$ be the respective projections onto the
eigenvectors $|\varphi_{\pm} \rangle$. Following (\ref
{diagonal})--(\ref{post-qubit}) 
we have a diagonal representation $\bM= \bQ_{+} - \bQ_{-}$; when we
measure observable $\bM$
on each qubit, the possible measurement results are $\pm1$; measuring
$\bM$ on the first qubit changes the state of the two-qubit system, and
after the measurement result $\pm1$ on the first qubit, the post-measurement
state of the two-qubit system is $\bQ_{\pm} \otimes\bI|\psi\rangle
/ \| \bQ_{\pm} \otimes\bI|\psi\rangle\|$.
Below we will evaluate the post-measurement state and show that
measuring $\bI\otimes\bM$ in the post-measurement state always
yields measurement results opposite to the measurement results on the
first qubit.

Since $(|0\rangle, |1\rangle)$ and $(|\varphi_+ \rangle, |\varphi
_-\rangle)$ are two bases for the one-qubit system in $\mathbb{C}^2$, then
%
\[
\left(
\matrix{ |0\rangle\cr |1\rangle
}
\right) =
\left[
\matrix{ \alpha_{11} & \alpha_{12} \cr \alpha_{21} & \alpha
_{22}
}
\right]
\left(
\matrix{ |\varphi_+ \rangle\cr |\varphi_- \rangle
}
\right),
\]
where $(\alpha_{j\ell})$ forms a $2 \times2$ unitary matrix with
determinant 
equal to a phase factor $e^{i\theta}$ ($i=\sqrt{-1}$) for some real
$\theta$.
Substituting the above expressions into the entangled state and
ignoring a global phase factor $e^{i\theta}$ (which has no effects
on measurement results; see Section~\ref{sec3.1}), we obtain
%
\begin{eqnarray} \label{spin}
|\psi\rangle&=& \frac{|01\rangle- |10\rangle}{\sqrt{2}} 
= e^{i\theta} \frac{|\varphi_+ \varphi_- \rangle- |\varphi_-
\varphi_+ \rangle}{\sqrt{2}}\nonumber
\\[-8pt]
\\[-8pt] &\sim&\frac{|\varphi_+ \varphi_-
\rangle- |\varphi_- \varphi_+ \rangle}{\sqrt{2}}.
\nonumber
\end{eqnarray}
From the definitions of $|\varphi_{\pm} \rangle$ and $\bQ_{\pm}$,
$\bQ_+ \otimes\bI|\varphi_+ \varphi_- \rangle= |\varphi_+
\varphi_- \rangle$,
$\bQ_- \otimes\bI|\varphi_- \varphi_+ \rangle= - |\varphi_-
\varphi_+ \rangle$,
$\bQ_+\otimes\bI|\varphi_- \varphi_+ \rangle= 0$, and
$\bQ_- \otimes\bI|\varphi_+ \varphi_- \rangle= 0$.
If the measurement result of $\bM$ on the first qubit is $+1$ (or
$-1$), from (\ref{spin}) we obtain the post-measurement state of the
two-qubit system as follows:
\begin{eqnarray*}
\frac{\bQ_{+} \otimes\bI|\psi\rangle}{\| \bQ_{+} \otimes\bI
|\psi\rangle\|}
& =& e^{i\theta}  \frac{\bQ_+ \otimes\bI
|\varphi_+ \varphi_- \rangle- \bQ_+ \otimes\bI|\varphi_- \varphi
_+ \rangle}{\sqrt{2} \| \bQ_{+} \otimes\bI|\psi\rangle\|} \\
&  =& e^{i\theta} \frac{|\varphi_+ \varphi_- \rangle}{ \| |\varphi
_+ \varphi_- \rangle\|} = e^{i\theta} |\varphi_+ \varphi_- \rangle
\sim|\varphi_+ \varphi_- \rangle
\end{eqnarray*}
(or $\bQ_{-} \otimes\bI|\psi\rangle/\| \bQ_{-} \otimes\bI|\psi
\rangle\| = e^{i\theta} |\varphi_- \varphi_+ \rangle\sim|\varphi
_- \varphi_+ \rangle$).
Since
\begin{eqnarray*}
\bI\otimes\bM|\varphi_+ \varphi_- \rangle&=& \bI\otimes\bQ_+
|\varphi_+ \varphi_- \rangle- \bI\otimes\bQ_- |\varphi_+ \varphi
_- \rangle\\
&=& - |\varphi_+ \varphi_- \rangle,\\
 \bI\otimes\bM|\varphi_- \varphi_+ \rangle&=& |\varphi_-
\varphi_+ \rangle,
\end{eqnarray*}
that is, the post-measurement state $|\varphi_+ \varphi_- \rangle$
(or\break $|\varphi_- \varphi_+ \rangle$) is the eigenvector of
$\bI\otimes\bM$ corresponding to eigenvalue $-1$ (or $+1$),
performing measurement on $\bI\otimes\bM$ in the post-measurement
state must always yield measurement result $-1$ (or $+1$). Thus,
the measurement results of $\bM$ on the two qubits of $|\psi\rangle$
are always opposite to each other.

\subsubsection{Quantum teleportation}\label{sec3.3.2}

Quantum teleportation is a process by which we can transfer the state of
a qubit from one location to another, without transmitting it
through the intervening space. We illustrate the phenomenon as follows.
Alice and Bob together generated a Bell state long ago. Each took one
qubit of the Bell state when they split.
Now they are far away from each other.
The mission for Alice is to deliver a qubit $|\psi\rangle$ to Bob,
while he is
hiding, and she can only send classical information to Bob but does not
know the state of the qubit $|\psi\rangle$.
Quantum teleportation is a way that Alice utilizes the entangled Bell
state to send a qubit of unknown state to Bob,
with only a small overhead of classical communication.
Recently a breakthrough in quantum teleportation has been made by
successfully transferring complex quantum data instantaneously
from one place to another, paving the way for real-world applications
of quantum communications (\cite*{Leeetal11}).

Here is how it works. Alice interacts the qubit $|\psi\rangle$ to be
teleported with her half of the Bell state, and
then performs a measurement on the two interacted qubits to obtain one
of four possible two-classical-bit results:
$00, 01, 10$ and $11$. She sends the two-bit information via classical
communication to Bob. Depending on Alice's
classical message, Bob performs one of four operations on his half of
the Bell state. Surprisingly,
the described procedure allows Bob to recover the original state $|\psi
\rangle$.

Specifically assume that the state to be teleported is $|\psi\rangle=
\alpha_0 |0\rangle+ \alpha_1 |1\rangle$, where $\alpha_0$ and
$\alpha_1$ are unknown amplitudes. First, 
consider a three-qubit state
\begin{eqnarray*}
|\varphi_0 \rangle&=& |\psi\rangle\frac{|00\rangle+ |11\rangle
}{\sqrt{2}}\\ &=& \frac{1}{\sqrt{2}}
 [ \alpha_0 |0 \rangle(|00\rangle+ |11\rangle) + \alpha_1 |1
\rangle(|00\rangle+ |11\rangle) ],
\end{eqnarray*}
where the first two qubits (on the left) belong to Alice, and the third
qubit to Bob. Note that Alice's second
qubit and Bob's third qubit are from the entangled Bell state. Second,
Alice applies a control-NOT gate to her qubits
in $|\varphi_0 \rangle$ and obtains
\[
|\varphi_1 \rangle= \tfrac{1}{\sqrt{2}}
 [ \alpha_0 |0 \rangle(|00\rangle+ |11\rangle) + \alpha_1 |1
\rangle(|10\rangle+ |01\rangle) ].
\]
Third, she applies a Hadamard gate to the first qubit in $|\varphi_1
\rangle$ and gets
\begin{eqnarray*}
|\varphi_2 \rangle&=& \tfrac{1}{2}
 [ \alpha_0 (|0 \rangle+ |1 \rangle) (|00\rangle+ |11\rangle)
\\
&&\hphantom{\tfrac{1}{2}
 [}{}+ \alpha_1 (|0 \rangle- |1 \rangle)(|10\rangle+ |01\rangle) ].
\end{eqnarray*}
We regroup the terms of $|\varphi_2 \rangle$ and rewrite it as follows:
%
\begin{eqnarray*}
|\varphi_2 \rangle&=& \tfrac{1}{2}
 [ |00\rangle(\alpha_0 |0 \rangle+ \alpha_1 |1 \rangle) +
|01\rangle(\alpha_0 |1 \rangle+ \alpha_1 |0 \rangle)\\
&&\hphantom{\tfrac{1}{2}
 [}{}+ |10\rangle(\alpha_0 |0 \rangle- \alpha_1 |1 \rangle)\\
 &&\hphantom{\tfrac{1}{2}
 [}{} +
|11\rangle(\alpha_0 |1 \rangle- \alpha_1 |0 \rangle)  ].
\end{eqnarray*}
The new expression has four terms, and each term has Alice's qubits in
one of four possible states
$|00\rangle,\break |01\rangle,  |10\rangle$ and $|11\rangle$, and Bob's
qubit is in the state related to
the original state $|\psi\rangle$. If Alice performs a measurement on
her qubits and informs Bob of the
measurement result, then his post-measurement state is completely
determined. For example, the first term has Alice's qubits in the state
$|00\rangle$ and
Bob's qubit in state $|\psi\rangle$. Therefore, if Alice's
measurement result on her qubits is $00$, then Bob's qubit will be in
state $|\psi\rangle$. Below is a list of Bob's four post-measurement
states corresponding to the results of Alice's measurements:
%
\begin{eqnarray*}
00 &\rightarrow&
\alpha_0 |0 \rangle+ \alpha_1 |1 \rangle, \quad
01 \rightarrow
\alpha_0 |1 \rangle+ \alpha_1 |0 \rangle,
\\
10 &\rightarrow&
\alpha_0 |0 \rangle- \alpha_1 |1 \rangle, \quad
11 \rightarrow
\alpha_0 |1 \rangle- \alpha_1 |0 \rangle.
\end{eqnarray*}
As Alice's measurement outcome on her qubits is one of $00, 01, 10$ and
$11$, depending on her measurement outcome
Bob's qubit will be one of the above four possible states. Once Alice
sends to Bob her two-classical-bit measurement outcome
through\break a~classical channel, he applies appropriate quantum gates to
his state and recovers
$|\psi\rangle$. For example, if her measurement is $00$, Bob's state
is $|\psi\rangle$, and he does not need to apply any quantum gate.
If her measurement is $01$, then Bob needs to apply a $\bsigma_x$ gate
to his state $\alpha_0 |1 \rangle+ \alpha_1 |0 \rangle$
and yields $|\psi\rangle$. If her measurement is $10$, then applying
a $\bsigma_z$ gate to his state $\alpha_0 |0 \rangle- \alpha_1 |1
\rangle$
Bob recovers $|\psi\rangle$. If her measurement is $11$, then Bob can
fix up his state
$\alpha_0 |1 \rangle- \alpha_1 |0 \rangle$ to recover $|\psi
\rangle$ by applying first a $\bsigma_x$ gate and then a\vadjust{\goodbreak}
$\bsigma_z$ gate. Here the $\bsigma_x$ and $\bsigma_z$ gates are
defined by Pauli matrices $\bsigma_x$ and $\bsigma_z$ given by~(\ref{Pauli}).
In summary, according to Alice's measurement outcome, applying some
appropriate quantum gates to his qubit Bob will recover the state
$|\psi\rangle$.

A few important remarks about quantum teleportation are in the line.
First, quantum teleportation does not involve any transfer of matter or
energy. Alice's particle has not been
physically moved to Bob; only its state has been transferred. Second,
after the teleportation Bob's qubit will be on the
teleported state, while Alice's qubit will become some undefined part
of an entangled state. In other words, what
the teleportation does is that a qubit was destroyed in one place but
instantaneously resurrected in another. 
Teleportation does not copy any qubits, and hence is consistent with
the no-cloning theorem (which forbids the creation of
identical copies of an arbitrary unknown quantum state; see \cite*
{WooZur82}). Third, in order to teleportate a qubit,
Alice has to inform Bob of her measurement by sending him two classical
bits of information.
These two classical bits do not carry complete information about the
qubit being teleported. If the two bits are intercepted by
an eavesdropper, he or she may know exactly what Bob needs to do in
order to recover the desired state.
However, this information is useless if the eavesdropper cannot
interact with the entangled particle in Bob's possession.
Also the requirement of sending two bits of information via classical
channel prevents quantum teleportation from transmitting
information faster than the speed of light.

\subsubsection{Bell's inequality}\label{sec3.3.3}
The Bell test experiments are designed to investigate
the validity of the entanglement effect in quantum mechanics through
Bell's inequality.
Over the past four decades many physical experiments on quantum systems
were conducted to check the validity of Bell's inequality and resulted
in some
violation of the inequality. For example,  Aspect, Grangier
and Roger (\citeyear{AspGraRog81,AspGraRog82N1,AspGraRog82N2})
provided overwhelming support to the violation of Bell's
inequality. The experimental results are often invoked as the proof of
quantum non-locality and lack of realism that no particle has definite
form until it is measured and 
measuring a quantum entity can instantaneously influence another far away.
See  Aspect, Grangier and Roger (\citeyear{AspGraRog81,AspGraRog82N1,AspGraRog82N2}),
 \citet{Boh51}, \citet
{Bel64}, \citet{Claetal69} and \citet{EinPodRos35}.
Below we describe the CHSH version of the Bell's inequality
(\cite*{Claetal69}).

Suppose $X_i$, $i=1,2,3,4$, are four random variables taking values
$\pm1$. Consider an ordinary experiment with two people, Alice and
Bob. In the experiment Alice observes $X_1$ or $X_2$ while Bob
meas\-ures~$X_3$ or $X_4$. Consider the quantity
$X_1 X_3 + X_2 X_3 + X_2 X_4 - X_1 X_4$. It is equal to
\[
(X_1 + X_2) X_3 + (X_2 - X_1) X_4 = \pm2 \leq2.
\]
Regardless of the distributions of $X_i$, taking expectation on both
sides of the above inequality we arrive at the famous Bell inequality,
%
\begin{eqnarray} \label{Bellinequality}
&&E(X_1 X_3) + E(X_2 X_3) + E(X_2 X_4)\nonumber
\\[-8pt]
\\[-8pt] && \quad {}- E(X_1 X_4) \leq2.
\nonumber
\end{eqnarray}

The violation of Bell's inequality demonstrates entanglement effect in
quantum mechanics. In fact, quantum experiments yield a
quantum version of the inequality. 
Consider that a quantum system of two qubits is prepared in a Bell state
\[
|\psi\rangle= \frac{|01\rangle- |10\rangle}{\sqrt{2}}.
\]
Alice takes the first qubit of $|\psi\rangle$ while Bob gets its
second qubit. Define four observables with eigenvalues $\pm1$,
\[
\bX_1 = \bsigma_z,  \quad  \bX_2 = \bsigma_x,
\]
on the first qubit and
\[
\bX_3 = - \frac{\bsigma_z + \bsigma_x}{\sqrt{2}},  \quad  \bX_4 =
\frac{\bsigma_z-\bsigma_x}{\sqrt{2}},
\]
on the second qubit, where $\bsigma_x$ and $\bsigma_z$ are Pauli
matrices given by (\ref{Pauli}).
Again Alice performs measurements on $\bX_1$ or $\bX_2$ while Bob
measures $\bX_3$ or $\bX_4$. The quantum expectations
of $\bX_1 \bX_3$, $\bX_2 \bX_3$, $\bX_2 \bX_4$, $\bX_1 \bX
_4$ in the state $|\psi\rangle$ are calculated below:
%
\begin{eqnarray*}
\mathbb{E}_\psi(\bX_1 \bX_3) &=&\frac{1}{\sqrt{2}},    \quad   \mathbb
{E}_\psi(\bX_2\bX_3)=\frac{1}{\sqrt{2}}, \\
  \mathbb{E}_\psi
(\bX_2\bX_4)&=&\frac{1}{\sqrt{2}},  \quad
\mathbb{E}_\psi(\bX_1\bX_4) = -\frac{1}{\sqrt{2}}.
\end{eqnarray*}
Here the observable product is in the sense of tensor product.
Thus we obtain a value in the quantum framework for the analog quantity
on the left-hand side of the Bell's inequality (\ref{Bellinequality})
\begin{eqnarray*}
&&\mathbb{E}_\psi(\bX_1\bX_3) + \mathbb{E}_\psi( \bX_2\bX_3)
+ \mathbb{E}_\psi( \bX_2\bX_4)\\
&& \quad {} - \mathbb{E}_\psi( \bX_1\bX
_4 ) = 2 \sqrt{2},
\end{eqnarray*}
which exceeds $2$ and hence violates the Bell's inequality.
In fact, the quantum version of the Bell's inequality is the
Tsirelson's inequality (Tsirelson, \citeyear{T80}) which shows that
in any quantum state $\brho$,
%
\begin{eqnarray} \label{Tsirelsoninequality}
&&\mathbb{E}_{\brho}(\bX_1\bX_3) + \mathbb{E}_{\brho}( \bX_2
\bX_3) + \mathbb{E}_{\brho}( \bX_2\bX_4) \nonumber
\\[-8pt]
\\[-8pt]
&& \quad {}- \mathbb{E}_{\brho}(
\bX_1\bX_4 ) \leq2 \sqrt{2}.
\nonumber
\end{eqnarray}

\subsection{Quantum Parallelism}\label{sec3.4}
Quantum computation has an amazing feature\linebreak[4]  termed as quantum
parallelism, which may be heu\-ristically
explained by the following oversimplifying description: a quantum
computer can simultaneously evaluate
the whole range of a function $f(x)$ at many different values of $x$.

For function $f(x)$ with $b$ bit input $x=x_1\cdots x_b$ and $1$ bit
output $f(x)$, we illustrate
quantum parallel evaluation of its values at many different $x$
simultaneously as follows.
First we apply $b$ Hadamard gates to the first $b$ qubits of $|0 \cdots
0 \rangle|0\rangle$ to obtain
\begin{eqnarray*}
&&\frac{|0 \rangle+ |1 \rangle}{\sqrt{2}} \cdots\frac{|0 \rangle+
|1 \rangle}{\sqrt{2}} |0\rangle
= \frac{1}{\sqrt{2^b}} \sum_x |x\rangle|0\rangle, \\
&& \quad
x=x_1\cdots x_b,    x_j=0,1,
\end{eqnarray*}
where the sum is over all possible $2^b$ values of $x$.
Second, apply quantum circuit $\bU_f$ defined in (\ref{Uf}) to the
obtained $b+1$ qubit state
to yield
\[
\frac{1}{\sqrt{2^b}} \sum_x |x \rangle |f(x) \rangle.
\]
The quantum circuit with $b$ Hadamard gates is extremely efficient in
producing an equal superposition of all
$2^b$ computational basis states with only $b$ gates; and quantum
parallelism enables simultaneous evaluation
of the whole range of the function $f$, although we evidently evaluate
$f$ just once with single quantum circuit
$\bU_f$ applied to the superposition state. 
To make it more clear we consider the case of $b=1$. Apply circuit $\bU_f$
to a superposition state as follows:
\begin{eqnarray*}
\bU_f \biggl( \frac{|0 \rangle+ |1 \rangle}{\sqrt{2}} | 0\rangle
 \biggr) &=& \bU_f \biggl( \frac{|00 \rangle+ |1 0\rangle}{\sqrt{2}}
 \biggr)\\
&=& \frac{|0 f(0) \rangle+ |1 f(1) \rangle}{\sqrt{2}}.
\end{eqnarray*}
One application of a single circuit $\bU_f$ results in a~superposition
state whose two components contain information about both
$f(0)$ and $f(1)$, as if we have evaluated $f(x)$ at values $0$ and $1$
simultaneously.
The quantum parallelism is in contrast with classical parallelism,
where multiple circuits each built to compute one value of
$f(x)$ are executed simultaneously. Quantum parallelism arises from
superposition states. A superposition state\vadjust{\goodbreak} has many components,
each of which may be thought of as a single argument to function
$f(x)$. Because of quantum nature, a single circuit $\bU_f$ applied
once to the superposition state is actually performed on each of the
components of the superposition, and the whole range of the values
of function $f(x)$ is stored in the resulted outcome superposition state.

The quantum parallelism can be a potentially powerful tool for
computational statistics. For example,
Bayesian analysis often encounters the problems of evaluating sums over
$2^b$ quantities, with $b$ proportional
to sample size or the number of variables. For moderate to large $b$,
the evaluation of such sums is computationally
prohibitive by classical computers 
(\cite*{Vid99}). Because of the quantum parallelism, it is
possible for quantum computers to perform such computing tasks.

\section{Quantum Algorithms}\label{sec4}

Quantum algorithms are described by quantum circuits that take input
qubits and
yield output measurements for the solutions of the given problems. As a
classical algorithm is a step-by-step
problem-solving procedure, with each step performed on a~classical
computer, a quantum algorithm is a step-by-step
procedure to solve a problem, with each step executed by a quantum
computer. Although all classical algorithms can
also be carried out on a quantum computer, we refer to quantum
algorithms as the algorithms that utilize essential
quantum features such as quantum superposition and quantum
entanglement. While it is true that
all problems solvable on a quantum computer are solvable on a classical
computer, and problems undecidable by
classical computers remain undecidable on quantum computers, what makes
quantum algorithms exciting is the faster
speed that quantum algorithms might be able to achieve, compared to
classical algorithms, for solving some tough problems.
The well-known quantum algorithms are Shor's factoring algorithm and
Grover's search algorithm. Shor's algorithm
and Grover's algorithm run, respectively, exponentially faster and
quadratically faster than the best known classical
algorithms for the same tasks.
Common techniques used in quantum algorithms include quantum Fourier
transform, phase estimation and quantum walk.

\subsection{Quantum Fourier Transform}\label{sec4.1}
The quantum Fourier transform is defined to be a linear transformation
on $n$ qubits that
maps the computational basis\vadjust{\goodbreak} states $|j\rangle$, $j=0, 1, \ldots,
2^n-1$, to superposition states
as follows:
\[
|j\rangle\longrightarrow\frac{1}{\sqrt{2^n}} \sum_{k=0}^{2^n-1}
e^{2\pi ijk/2^n}|k\rangle, \quad i =\sqrt{-1}.
\]
The inverse of quantum Fourier transform is given by
\[
|k\rangle\longrightarrow\frac{1}{\sqrt{2^n}} \sum_{j=0}^{2^n-1}
e^{-2\pi ijk/2^n}|j\rangle.
\]

We use the binary representation to express the state $j=j_12^{n-1} +
j_22^{n-2} + \cdots+ j_n2^0$
as $j=j_1j_2\cdots j_n$ and represent binary fraction $j_\ell/2 +
j_{\ell+1}/2^2 + \cdots+ j_m/2^{m-\ell+1}$ as
$0.j_{\ell}j_{\ell+1}\cdots j_m$, where $1\leq\ell\leq m \leq2^n$.
Then the quantum Fourier transform of state $|j\rangle= |j_1j_2\cdots
j_n\rangle$ has the following useful
product representation:
\begin{eqnarray*}
|j_1j_2\cdots j_n\rangle&\rightarrow&\frac{1}{2^{n/2}}
 ( |0\rangle+ e^{2\pi i0.j_n} |1\rangle )\\
 &&{}\cdot
 ( |0\rangle+ e^{2 \pi i 0.j_{n-1}j_n} |1\rangle )\cdots\\
&&{}\cdot
 ( |0\rangle+ e^{2 \pi i 0.j_1j_2\cdots j_n} |1\rangle
).
\end{eqnarray*}
It can be easily checked from the product representation that 
with quantum parallelism the quantum Fourier transform can be realized
as a quantum circuit with only $O(n^2)$ operations,
while classically the fast Fourier transform requires $O(n 2^n)$
operations for processing $2^n$ data,
which indicates an exponential speed-up (\cite*{NieChu00}).
Realizing such an exponential saving accommodated by quantum
parallelism requires clever measurement
schemes. Successful examples include 
quantum phase estimation and Shor's algorithms for factoring and
discrete logarithm.

\subsection{Phase Estimation}\label{sec4.2}
Quantum algorithms are of random nature in the sense that they are able
to produce correct answers
only with some probabilities. Consider quantum\break phase estimation which
provides the key to many quantum
algorithms. Assume that a unitary operator $\bU$
has an eigenvector $|x \rangle$ with eigenvalue $e^{2 \pi i
\varphi}$. The phase~$\varphi$ of
the eigenvalue is unknown and the goal of the phase estimation
algorithm is to estimate
$\varphi$ based on the assumption that the state $|x \rangle$ can be
prepared and
the controlled-$\bU^{2^j}$ operations [see Section~\ref{sec3.2} for control
gate] can be performed for suitable nonnegative integers $j$.

The registers are used in phase estimation. The first register consists of
$b$ qubits initially in the state $|0\rangle$.
The second register starts\vadjust{\goodbreak} in the state $|x\rangle$ and involves
enough qubits to store $|x \rangle$.
The phase estimation procedure is performed in two stages.

First, we apply Hadamard transform to the
first register and then controlled-$\bU$ operations on the second
register, with $\bU$ raised to successive
powers of 2, to obtain the final state with the second register
unchanged and the first register given by
%
\begin{eqnarray} \label{PE1}\qquad
&&\frac{1}{2^{b/2}}  ( |0\rangle+ e^{2 \pi i 2^{b-1} \varphi
} | 1 \rangle )\nonumber\\
&& \qquad {}\cdot
 ( |0\rangle+ e^{2 \pi i 2^{b-2} \varphi} | 1 \rangle
 ) \cdots
 ( |0\rangle+ e^{2 \pi i 2^{0} \varphi} | 1 \rangle
) \\
&& \quad =
\frac{1}{2^{b/2}} \sum_{k=0}^{2^b-1} e^{2 \pi i \varphi k} | k
\rangle.\nonumber
\end{eqnarray}
If $\varphi$ is expressed exactly in $b$ bits as $\varphi=0.\varphi
_1\cdots\varphi_b$, (\ref{PE1}) becomes
%
\begin{eqnarray} \label{PE2}
&&\frac{1}{2^{b/2}}  ( |0\rangle+ e^{2 \pi i 0.\varphi_b} | 1
\rangle )\nonumber
\\
 && \quad {}\cdot( |0\rangle+ e^{2 \pi i 0.\varphi_{b-1}\varphi_b} | 1
\rangle ) \cdots\\
 && \quad {}\cdot
 ( |0\rangle+ e^{2 \pi i 0.\varphi_1\varphi_1\cdots\varphi
_b} | 1 \rangle ),
\nonumber
\end{eqnarray}
which is the quantum Fourier transform of the product state $|\varphi
_1\varphi_2\cdots\varphi_b\rangle$.

The second stage of phase estimation is to take the inverse quantum
Fourier transform on the
first register. For $\varphi=0.\varphi_1\cdots\varphi_b$, the
output state from the second stage is $|\varphi_1\varphi_2\cdots
\varphi_b\rangle$, and a measurement in the computational basis yields
$\varphi_1\cdots\varphi_b$ and dividing the measurement by $2^b$ gives
$\varphi_1\cdots\varphi_b/2^b=0.\varphi_1\cdots\varphi_b=\varphi
$. We obtain a perfect estimate of $\varphi$.

Now we consider the case that $\varphi$ cannot be expressed exactly
with a $b$ bit binary expansion.
Take $0 \leq\eta< 2^b$ to be the integer that its binary fraction
$\eta/2^b=0.\eta_1\eta_2\cdots\eta_b$
is the first $b$ bit representation in the binary expansion of $\varphi
$, which satisfies
$0 \leq\varphi-\eta/2^b \leq2^{-b}$.

Perform the inverse quantum Fourier transform on the first register
given by (\ref{PE1}), which is obtained
in the first stage results, and get
\[
\frac{1}{2^{b}} \sum_{k,\ell=0}^{2^b-1} e^{-2 \pi i k \ell
/2^b}  e^{2 \pi i \varphi k} | \ell\rangle
= \sum_{\ell=0}^{2^b-1} \beta_\ell |\ell\rangle,
\]
where amplitudes of $|(\eta+ \ell) (\operatorname{mod}  2^b)
\rangle$ are
%
\begin{eqnarray*}
\beta_\ell&=& \frac{1}{2^{b}} \sum_{k=0}^{2^b-1}  \bigl\{e^{2 \pi
i [\varphi- (\eta+\ell)/2^b]}  \bigr\}^k\\
&=& \frac{1}{2^b} \biggl ( \frac{1- e^{2 \pi i (2^b \varphi- \eta
- \ell)} }{1- e^{2 \pi i (\varphi- \eta/2^b - \ell/2^b)} }
 \biggr).
\end{eqnarray*}
Assume that the result of the final measurement from phase estimation
is $\tilde{\eta}$ and dividing the result by
$2^b$ gives $\tilde{\varphi}=\tilde{\eta}/2^b$. Let $\zeta$ be the
specified accuracy for the phase estimation procedure.
By adding up $|\beta_\ell|^2$ with $\ell$ being within $\zeta
2^b$, we bound the probability that the obtained
$\tilde{\varphi}$ is within $\zeta$ from $\varphi$:
\begin{eqnarray*}
P(|\tilde{\varphi}-\varphi| \leq\zeta) &\geq&
P (|\tilde{\eta} - \eta| \leq\zeta 2^b -1  ) \\
&\geq&1 -
\frac{1}{2 (\zeta 2^b-2)}.
\end{eqnarray*}
For $\epsilon>0$, set
%
\begin{equation} \label{b}
b =  \biggl[\log_2 \biggl( \frac{1}{\zeta}  \biggr)  \biggr] + \biggl [
\log_2  \biggl(2 + \frac{1}{2 \epsilon}  \biggr)  \biggr].
\end{equation}
Then $P(|\tilde{\varphi}-\varphi| \leq\zeta) \geq1 - \epsilon$,
that is, with probability at least $1-\epsilon$
the phase estimation procedure can successfully produce $\tilde
{\varphi}$ within $\zeta$ from the true $\varphi$.
See \citet{NieChu00}.

\subsection{Statistical Analysis}\label{sec4.3}

The phase estimation algorithm requires $b$ qubits for the first
register to achieve $[-\log_2 \zeta]$ bit accuracy
and success probability $1-\epsilon$. With accuracy fixed, to increase
the success probability the required qu\-bits
\[
b \sim1-\log_2 \zeta+ \frac{1}{4 \epsilon \log 2},
\]
which grows at a very fast rate. For example, an increase in success
probability from $90\%$ to $99\%$ requires eighteen times of
qubit increase compared to the change from $80\%$ to $90\%$.

Quantum algorithms are of random nature in the sense that they often
produce correct answers only with certain
probabilities. The success probabilities depend upon the schemes of the
algorithms as well as the context of applications.
Given a quantum algorithm for solving a problem, a common practice is
to repeatedly run the quantum algorithm to
achieve high probability of successfully obtaining a~correct answer.
Consider that the phase estimation
procedure is repeatedly run $n$ times to obtain results $\tilde
{\varphi}_1, \ldots, \tilde{\varphi}_n$. Then
$\tilde{\varphi}_1, \ldots, \tilde{\varphi}_n$ may be treated as
i.i.d. random variables with each $\tilde{\varphi}_j$ satisfying
\[
P( |\tilde{\varphi}_j - \varphi| > \zeta) \leq\epsilon.
\]
We may statistically model $\tilde{\varphi}_j$ by the gross error
model (\cite{HubRon09}) as follows.
Assume that $\tilde{\varphi}_j$ are independently and identically
generated from $(1-\epsilon) F(x) + \epsilon H(x)$, where
$F(x)$ is the distribution of the correct answers that are within~$\zeta$ from true $\varphi$,
and $H(x)$ is the distribution of wrong answers\vadjust{\goodbreak} that are at least
$\zeta$ away from true $\varphi$. Then~$\tilde{\varphi}_j$ are
correct with probability $1-\epsilon$ and
incorrect with probability $\epsilon$.

If the outcome result of the algorithm is verifiable to be a correct
answer or not
[as in the case of Shor's algorithms for factoring and order-finding in
Section~\ref{sec4.4} below],
the obtained result from each run is checked to be a correct answer or not.
Then the number of times required to run the algorithm for obtaining a
correct answer follows a geometric distribution. Thus the
probability that we obtain a~correct answer in $n$ repetitions is equal to
\begin{eqnarray*}
&&P(\mbox{obtain a correct answer in $n$ trials}) \\
&& \quad = 1 - P(\mbox{no
success in the $n$ trials}) = 1 - \epsilon^n.
\end{eqnarray*}
Since $\epsilon^n$ goes to zero geometrically fast,
we may choose a moderate $\epsilon$ with fewer qubits to achieve very
high probability of successfully obtaining
a correct answer by repeatedly running the algorithm enough times.

On the other hand, if the outcome result is not verifiable to be a
correct answer or not
[as in the case of phase estimation], careful analysis is needed
to design ways for obtaining a correct answer with very high
probability. As wrong answers are far away
from true $\varphi$, estimators like sample average of $\tilde
{\varphi}_1, \ldots,\break \tilde{\varphi}_n$ may not estimate
$\varphi$ well. We adopt a robust statistical method to estimate
$\varphi$ by $\alpha$-trimmed mean~$\bar{\varphi}$,
which is defined as follows. Ordering $\tilde{\varphi}_1, \ldots,
\tilde{\varphi}_n$
and then removing $[n \alpha]$ largest and $[n \alpha]$ smallest
ones, we take the average of the remaining $\tilde{\varphi}_j$
as $\alpha$-trimmed mean, where $\alpha$ is chosen to be greater than
$\epsilon/2$. One example is the sample
median of $\tilde{\varphi}_1, \ldots, \tilde{\varphi}_n$. The
probability that $\bar{\varphi}$ is within
$\zeta$ from $\varphi$ can be calculated from the binomial
probability as follows:
\begin{eqnarray*}
  P( |\bar{\varphi} - \varphi| \leq\zeta) &\geq& P\bigl(\mbox{more than
$n (1 - 2 \alpha)$ number}\\
&& \hspace*{16pt}\hphantom{P\bigl(}\mbox{of $\tilde{\varphi}_j$ are within
$\zeta$ from $\varphi$}\bigr) \\
  &=& \sum_{k=[n (1-2 \alpha)]-1}^n \left(
\matrix{ n \cr k
}
\right)  (1-\epsilon)^k \epsilon^{n-k}.
\end{eqnarray*}
%
As $n \rightarrow\infty$, $n \epsilon$ approaches to infinity. The
binomial probability
can be approximated by resorting to a~normal approximation, yielding
\begin{eqnarray*}
&& \sum_{k=[n (1-2 \alpha)]-1}^n \left(
\matrix{ n \cr k
}
\right)  (1-\epsilon)^k \epsilon^{n-k} \\
&& \quad \sim
1 - \Phi \biggl( \frac{ \sqrt{n} (\epsilon- 2 \alpha)}{\sqrt
{\epsilon (1-\epsilon)}}  \biggr)
=\Phi \biggl( \frac{ \sqrt{n} (2 \alpha- \epsilon)}{\sqrt
{\epsilon (1-\epsilon)}}  \biggr),
\end{eqnarray*}
where $\Phi(\cdot)$ is the standard normal distribution function.
Since $2\alpha-\epsilon> 0$,\vadjust{\goodbreak} as $n$ increases, $P( |\bar{\varphi} -
\varphi| \leq\zeta)$ approaches to $1$ exponentially fast.
Combining the two cases together, we arrive at the following theorem.
\begin{thm}
Suppose that the outcome $\tilde{\varphi}$ of\break a~quantum algorithm
obeys the gross error model
that with probability $1-\epsilon$ it produces a correct answer and
probability $\epsilon$ it gives
a wrong answer. Then by repeatedly running the quantum algorithm we
will obtain a correct answer with probability
approaching 1 exponentially fast in the number of repetitive runs.
\end{thm}

For a quantum algorithm that produces a correct answer with probability
$70\%$ and $\alpha=0.2$, in order to obtain a correct answer
with $0.999$ probability we need to run the algorithm five times and 20
times, respectively, for the cases that the
outcome results are verifiable and not verifiable.


\subsection{Factoring and Order-Finding Algorithms}\label{sec4.4}
The factoring problem is to find all prime factors of a given positive
composite number such
that the product of these prime numbers is equal to the composite
number. Factoring is known to be a
very hard problem for classical computers.  Shor (\citeyear{Sho94,Sho97}) developed a quantum algorithm
for the factoring problem that is exponentially faster than the most
efficient known classical
factoring algorithm.

Shor's quantum algorithms work as follows. Mathematically
the factoring problem is equivalent to the order-finding problem that
for two positive integers~$x$ and $N$, $x < N$, with no common factors,
find the smallest integer $r$
such that dividing $x^r$ by $N$ we obtain a reminder $1$ (\cite*{Sho97};
\cite*{NieChu00}). The quantum algorithm for factoring is reduced
to a quantum algorithm for order-finding.
The quantum algorithm for order-finding is to apply the phase
estimation algorithm to the unitary operator
\[
\bU  |y \rangle= |x  y (\operatorname{mod} N) \rangle.
\]
The eigenvectors of $\bU$ are
\begin{eqnarray*}
&&|u_s\rangle= \frac{1}{\sqrt{r}} \sum_{k=0}^{r-1} \exp \biggl( \frac
{-2 \pi i s k}{r}  \biggr) |x^k \operatorname{mod } N \rangle,\\
&&
\quad s=0,1, \ldots, r-1,   i = \sqrt{-1},
\end{eqnarray*}
with corresponding eigenvalues $\exp(2 \pi i s/r)$. Using the
phase estimation algorithm we can obtain the eigenvalues
$\exp(2 \pi i s/r)$ with high accuracy and thus find the order $r$
with certain probability.

While the quantum factoring algorithm can accomplish the task of\vadjust{\goodbreak}
factoring an $n$-bit integer with operations of order $n^2 \log n
\log\log n$,
the current best known classical algorithm requires operations of order
$\exp(n^{1/3} \log^{2/3} n)$ to factor an $n$-bit composite number
(\cite*{CraPom01}). 
Note that the number of operations required in the best classical
algorithm grows exponentially in the size of the number being factored.
Because of the exponential complexity, the factoring problem is
generally regarded as an intractable problem on classical computers.

The factoring problem plays an important role in cryptography.
Cryptography is to enable two parties, Alice and Bob, to communicate privately,
while it is very difficult for the third parties to ``eavesdrop'' on the
contents of the
communications. Examples include ATM cards, computer passwords,
internet commences, clandestine meetings and military communications.
Two cryptographic protocols used in the communications are private key
cryptosystem and public key cryptosystem.
A private key cryptosystem requires the two communicating parties to
share a private key. Alice uses the key to encrypt the information,
sends the encrypted information to Bob who uses the key to decrypt the
received information. The severe drawback of
the private key cryptosystem is that the parties have to safeguard the
key transmission from being eavesdropped. A~public key
cryptosystem invented in the 1970s requires no sharing secret key
in advance. Bob publishes a ``public key'' available to the general
public, and Alice uses the public key to encrypt
information and sends the encrypted information to Bob. The encryption
transformation is specially created such that with only the public
key, it is extremely difficult, though not impossible, to invert the
encryption transformation. When publishing the public key Bob keeps a matched
secret key for easy inversion of the encryption transformation and
decryption of the received information.
One of the most widely used public key cryptosystems is the RSA cryptosystem,
which is named after its creator Rivest, Shamir and Adleman
(\cite*{MenvanVan96}; \cite*{RivShaAdl78}).
RSA is built on the mathematical asymmetry of factoring: it is easy to
multiply large prime numbers and obtain their product as a composite number
but hard to find the prime factors of a given large composite number.
RSA encryption keeps the large primes as a secret key and uses their product
to make a~``public key.'' Because of its exponential complexity,
tremendous efforts tried to break the RSA system so far have\vadjust{\goodbreak} resulted
in vain,
and there is a~widespread belief that the RSA system is secure against
any classical computer based attacks.
As the factoring problem can be efficiently solved by Shor's quantum
factoring algorithm, a quantum computer can break the RSA system easily.
Fortunately, while quantum mechanics takes away with one hand, it gives
back with the other. A quantum procedure known as quantum cryptography or
quantum key distribution can do key distribution so that the
communication security cannot be compromised. The idea is based on the
quantum principle
that observing a quantum system will disturb the system being observed.
If there is an eavesdropper during the transmission of the quantum key
between Alice and Bob,
eavesdropping will disturb the quantum communication channel that is
used to establish the key, and the disturbance will make eavesdropping visible.
Alice and Bob will throw away the compromised key and keep only the
secured key for their communication.

\subsection{Quantum Search Algorithm}\label{sec4.5}
Suppose that you would like to find the name corresponding to a given
phone number in a telephone directory; or suppose that there are some
locations in a given city you would like to visit and wish to find the
shortest route passing through all the locations. If there are $N$
names in the telephone directory or~$N$ possible routes to pass through
all the locations, search algorithms by classical computers usually
require operations of order~$N$. One such simple classical algorithm is
to check exhaustively all names to find a name matching with the given
phone number or to search all possible routes and then find the
shortest route among all routes. However,  Grover (\citeyear{Gro96,Gro97})  developed a quantum search algorithm that
needs only operations of order $\sqrt{N}$ to
find a solution to the search problem.

The quantum search algorithm works as follows. Suppose that the search
space has $N$ elements and the search
problem has exactly $M$ solutions. Assume $M \leq N/2$. (For the silly
case of $ M > N/2$, we either search for the solution by doing random selection
from the search space or double the number of the elements in the
search space by adding $N$ extra non-solution elements to the search space.)
The algorithm works by creating superposition state with Hadamard gate,
\[
| \psi\rangle= \frac{1}{N^{1/2}} \sum_{x=0}^{N-1} |x\rangle,
\]
and then applying a so-called Grover iteration (or operator) 
repeatedly. Set
\[
|\phi\rangle= \frac{1}{\sqrt{N-M}} \sum_{x^\prime} |x^\prime
\rangle, \quad|\varphi\rangle=
\frac{1}{\sqrt{M}} \sum_{x^{\prime\prime}} |x^{\prime\prime
}\rangle,
\]
where the summations over $x^\prime$ and $x^{\prime\prime}$ denote
sums over all non-solutions and solutions, respectively.\break
Then we can express $|\psi\rangle$ as follows:
\[
|\psi\rangle= \sqrt{\frac{N-M}{N}}|\phi\rangle+ \sqrt{\frac
{M}{N}} |\varphi\rangle.
\]
The Grover operator is to perform two reflections, one about the vector
$|\phi\rangle$ and another about
the vector $|\psi\rangle$. The two reflections together are a
rotation with angle $\theta$ in the two-dimensional space spanned by
$|\phi\rangle$ and $|\varphi\rangle$, where
\[
\cos(\theta/2) = \sqrt{ \frac{N-M}{N}}.
\]
After the rotation, the initial state $|\psi\rangle=\cos(\theta/2)
 |\phi\rangle+ \sin(\theta/2)  |\varphi\rangle$ becomes state
\[
\cos(3 \theta/2)  |\phi\rangle+ \sin(3 \theta/2)  |\varphi
\rangle.
\]
Thus each application of the Grover operator is a~rotation with angle
$\theta$.
The initial state $|\psi\rangle$ has angle $\pi/2 - \theta/2$ with
$|\varphi\rangle$; after the first rotation,
the resulted state has angle $\pi/2 - 3 \theta/2$ with $|\varphi
\rangle$; and in general after the $r$th rotation,
the resulted state has angle $\pi/2 - (2 r+1) \theta/2$ with
$|\varphi\rangle$.
Repeatedly applying the Grover operator, we rotate the state vector
near $|\varphi\rangle$.
With the initial state $|\psi\rangle=\cos(\theta/2)  |\phi\rangle
+ \sin(\theta/2)  |\varphi\rangle$,
we need to rotate through $\arccos\sqrt{M/N}$ radians to transform
the state vector to $|\varphi\rangle$.
After
$ R= \arccos( \sqrt{M/N})/\theta= O(\sqrt{N/M})
$
times of applications of the Grover operator, we rotate the state
vector $|\psi\rangle$ to within an angle $\theta/2$ of $|\varphi
\rangle$.
Performing measurements of the state yields a solution to the search
problem with probability at least $\cos^2(\theta/2)\geq1 - M/N$. %

The number of iterations $R$ depends on $M$, the number of solutions.
Since $R \leq\pi/(2 \theta)$ and $\theta/2 \geq\sin(\theta
/2)=\sqrt{M/N}$,
$R \leq(\pi/4) \sqrt{N/M}$. Typically,\break $M \ll N$, $\theta\approx
\sin\theta\approx2 \sqrt{M/N}$, thus $R \approx(\pi/4)\cdot\sqrt{N/M}$.
We estimate the number of solutions by quantum counting,
which is to combine the Grover operator with the phase estimation method.
Under the basis $|\phi\rangle$ and $|\varphi\rangle$ the Grover
operator has 
eigenvalues $e^{i \theta}$ and $e^{i (2\pi- \theta)}$. Applying the
phase estimation method we can estimate the
eigenvalues and thus~$\theta$ with prescribed precision and
probability, which in turn yields $M$.
The combination of the quantum counting and search procedure will find
a solution of the search problem with certain probability.
Repeating the quantum search algorithm will boost the probability and
enable us to obtain a solution to the search problem.

Quantum walk and quantum Markov chain are currently being investigated
for new quantum search algorithms
and quantum speed-up of Markov chain based algorithms (\cite*{AhaTaS03}, \cite*
{Chietal}; \cite*{Chi10}; \cite*{Tul08}; \cite*{SheKemWha03} and Szegedy, \citeyear{Sze}).
In Section~\ref{sec5} we show that the quantum search algorithm can also be
viewed as a quantum simulation procedure.

\section{Quantum Simulation}\label{sec5}
Quantum simulation is to intentionally and artificially mimic a natural
quantum dynamics, which is hard to access,
and analyze, by a computer-gener\-ated quantum system, which is easy to
manipulate and investigate.
It provides scientific means for simulating complex biological, chemical
or physical systems in order to study and understand certain scientific
phenomena and evaluate hard-to-obtain
quantities in the systems. Examples in modern scientific studies
include the estimation of dielectric constant, proton mass, and
precise energy of molecular hydrogen, the study of superconductivity,
the test of novel nano-materials, and the
design of new biomolecules.\looseness=1

To simulate a quantum system we need to solve the Schr\"odinger
equation (\ref{schrodinger1}) which governs the dynamic evolution of
the system.
For a typical Hamiltonian with real particles the Schr\"odinger
equation usually consists of
elliptical differential equations, each of which can be easily
simulated by a classical computer.
However, the real challenge in simulating a quantum system is to solve
the exponential
number of such differential equations. For a quantum system of $b$
qubits, its states have
$2^b$ amplitudes. To simulate the dynamic behavior of $b$ qubits
evolving according to
the Schr\"odinger equation, we need to solve a system of $2^b$
differential equations.
Due to the exponential growth in the number of differential equations,
the simulation of general quantum systems by classical computers is
very inefficient.
Classical simulation of quantum systems is feasible for the cases where
insightful approximations are available to dramatically reduce the
effective number of
equations involved.
Quantum computers may excel in simulating physically\vadjust{\goodbreak} interesting and
important quantum systems
for which efficient simulation by classical computers may not be available.

\subsection{Simulate a Quantum System}\label{sec5.1}

The key of quantum simulation is to solve the Schr\"odinger equation
(\ref{schrodinger1})
which has solution
%
\begin{equation} \label{schrodinger2}
| \psi(t) \rangle= e^{-i \bH t}   | \psi(0) \rangle, \quad i =
\sqrt{-1}.
\end{equation}
Numerical evaluation of $e^{-i \bH t}$ is needed.
The Hamiltonian $\bH$ is usually exponentially large and extremely
difficult to
exponentiate. The common approach in numerical analysis is to use the
first-order
linear approximation, $1 - i \bH \delta$, of $e^{-i \bH (t +
\delta)} -
e^{-i \bH t}$, which often yields unsatisfactory numerical solutions.

Many classes of Hamiltonians have sparse representations. For such
sparse Hamiltonians
we can find efficient evaluation of the solutions (\ref{schrodinger2})
with high-order approximation. For example, the Hamiltonians in many
physical systems involve
only local interactions, which originate from the fact that most
interactions fall off
with increasing distance in location or increasing difference in
energy. In the local Hamiltonian case, the Hamiltonian of
a quantum system with $\alpha$ particles in a $d$-dimensional space
has the form
%
\begin{equation} \label{Hamiltonian}
\bH= 2 \sum_{\ell=1}^L \bH_\ell,
\end{equation}
where $L$ is a polynomial in $\alpha+d$, and each $\bH_\ell$ acts on
a small subsystem of size
free from $\alpha$ and $d$. For example, the terms $\bH_\ell$ are typically
two-body interactions and one-body Hamiltonians.
Hence $e^{-i \bH_\ell \delta}$ are easy to approximate, although
$e^{-i \bH \delta}$ is very hard to evaluate. Since $\bH_\ell$
and $\bH_k$ are non-commuting,
$e^{-i \bH \delta} \neq e^{-i \bH_1 \delta} \cdots e^{-i \bH
_L \delta}$. Applying a
modification of the Trotter formula (\cite*{Kat78}; \cite*{Tro59}; \cite*{autokey82}) we obtain
\[
e^{-i \bH \delta} = \bU_{\delta} + O(\delta^2),
\]
where
%
\begin{equation} \label{eqU}
 \quad \bU_{\delta} =  [ e^{-i \bH_1 \delta} \cdots e^{-i \bH_L
\delta}  ]
 [ e^{-i \bH_L \delta} \cdots e^{-i \bH_1 \delta}  ].
\end{equation}
Thus we can approximate $e^{-i \bH \delta}$ by $\bU_{\delta}$ which
needs to evaluate only each $e^{-i \bH_\ell \delta}$.

Assume that the quantum system starts at $t=0$ with initial state
$|\psi(0)\rangle$
and ends at final time $t=1$. For an integer $m$, set $t_j=j/m$,
$j=0,1,\ldots, m$. The quantum
simulation is to apply approximation $\bU_\delta$ of $e^{-2 i \bH
\delta}$ to evaluate
(\ref{schrodinger2}) at $t_j$ iteratively and generate approximate
solutions to $|\psi(t_j)\rangle$.
Denote by $|\tilde{\psi}(t_j) \rangle$ the state at $t_j$ obtained
from the quantum simulation as\vadjust{\goodbreak}
an approximation of the true state $|\psi(t_j)\rangle$ at~$t_j$. Then
for $j=1,\ldots, m$,
%
\begin{eqnarray} \label{finalstate}
 \qquad |\psi(t_{j})\rangle&=& e^{-2 i \bH \delta}  |\psi
(t_{j-1})\rangle= e^{-2 i \bH j \delta}  |\psi(t_0)\rangle,
\nonumber
\\[-8pt]
\\[-8pt]
 \qquad |\tilde{\psi}(t_{j})\rangle&=& \bU_\delta |\tilde{\psi
}(t_{j-1})\rangle= \bU^j_\delta |\psi(t_0) \rangle.
\nonumber
\end{eqnarray}
While classical computers are inefficient in simulating general quantum
systems, quantum computers can efficiently
carry out the quantum simulation procedure and provide an exponential
speedup for the quantum simulation on classical
computers. In spite of the inefficiency, classical computers are
currently being used to simulate quantum systems in
biochemistry and material science. Quantum simulation will be among the
important applications of quantum computers.
See \citet{AbrLlo97}, \citet{AspDutHea}, \citet{Benetal02}, \citet{Beretal07}, \citet{FreKitWan02}, \citet{Janetal03}, \citet{BogTay98}, \citet{Llo96}, \citet
{NieChu00}, and \citet{Zal98}.

\subsection{Recast Quantum Search Algorithm as Quantum Simulation}\label{sec5.2}
Grover's search algorithm discussed in Section~\ref{sec4.5} is an important
finding in quantum computation. It can be heuristically
sketched as a quantum simulation by writing down an explicit
Hamiltonian $\bH$ such that
a quantum system evolves from its initial state $|\psi\rangle$ to
$|x\rangle$ after some specified time, where $x$ is a solution
of the search problem. Of course the Hamiltonian $\bH$ depends on the
initial state $|\psi\rangle$ and solution $x$. Suppose that
$|y \rangle$ is another state such that $|x\rangle$ and $|y\rangle$
form an computational basis, and
$|\psi\rangle= \alpha |x\rangle+ \beta |y\rangle$ for real
$\alpha$ and $\beta$ with $\alpha^2 + \beta^2=1$.
Define Hamiltonian
\[
\bH= |x \rangle\langle x| + | \psi\rangle\langle\psi| = \bI+
\alpha (\beta \bsigma_x + \alpha \bsigma_z),
\]
where $\bsigma_x$ and $\bsigma_z$ are Pauli matrices defined in (\ref
{Pauli}). Then
\begin{eqnarray*}
&&\exp(-i \bH t)  |\psi\rangle\\
&& \quad = e^{-i t}  [ \cos(\alpha
t)  |\psi\rangle- i  \sin(\alpha t)  (\beta \bsigma_x +
\alpha \bsigma_z)
 |\psi\rangle ]\\ && \quad = e^{-i t}  [ \cos(\alpha t)  |\psi
\rangle- i  \sin(\alpha t)  |x \rangle ].
\end{eqnarray*}
Measuring the system at time $t=\pi/(2 \alpha)$ yields the solution
state $|x\rangle$.

\subsection{Quantum Monte Carlo Simulation}\label{sec5.3}

Quantum theory is intrinsically stochastic and\break quantum measurement
outcome is random. As many naturally occurring
quantum systems involve a large number of interacting particles,
due to the computational complexity we\vadjust{\goodbreak} are forced to utilize Monte
Carlo techniques in the simulations of
such quantum systems. The combination of Monte Carlo methods with
quantum simulation makes it possible
to obtain reliable quantifications of quantum phenomena and estimates
of quantum quantities.
Such combination procedures are often referred to as quantum Monte
Carlo simulation (\cite*{autokey52}; \cite*{Rou08}).
Consider the problem of estimating the following quantity:
%
\begin{equation} \label{theta}
\theta= \operatorname{Tr}(\bX  \brho) = E(X),
\end{equation}
where $\bX$ is an observable, $X$ is its measurement result, and
$\brho$ is the state of the quantum
system under which we perform the measurements and evaluate the
quantity $\theta$.

As $\brho$ is the true final state of the quantum system, we denote by
$\tilde{\brho}$ the final state of
the quantum system obtained via quantum simulation.
The quantum systems are prepared in initial state $\brho_0$, 
and we use the quantum simulation procedure described above to simulate
the evolutions of the systems from initial state $\brho_0$ 
to final state $\tilde{\brho}$ 
according to Schr\"odinger's equation
(\ref{schrodinger2}) with Hamiltonian $H$ given by (\ref
{Hamiltonian}). We repeatedly perform the measurements
of such $n$ identically simulated quantum systems at the final state
and obtain measurement
results $X_1,\ldots, X_n$. We estimate $\theta$ defined in (\ref
{theta}) by
%
\begin{equation} \label{thetahat}
\hat{\theta} = \frac{1}{n} \sum_{j=1}^n X_j.
\end{equation}

The target $\theta$ given by (\ref{theta}) is defined under the true
state $\brho$,
while the simulated quantum system is under approximate final state
$\tilde{\brho}$ which is close to $\brho$.
The measurement results $X_1, \ldots, X_n$ are obtained via quantum
simulation from the quantum systems
in the simulated state $\tilde{\brho}$. Therefore, the Monte Carlo
quantum estimator $\hat{\theta}$
in (\ref{thetahat}) involves both bias and variance. \citet
{Wan11} studied the quantum simulation procedure
and investigated the bias and variance of $\hat{\theta}$. The derived
bias and variance results can be used
to design optimal strategy for the best utilization of computational
resources to obtain
the quantum Monte Carlo estimator.
\section{Quantum Information}\label{sec6}

Classical information theory is centered on Shannon's two coding
theorems on noiseless and noisy channels. The noiseless channel coding
theorem quantifies the number of classical bits required to store
information for transmission by Shannon entropy, while the
noisy\vadjust{\goodbreak}
channel coding theorem quantifies the amount of information that can be
reliably transmitted through a noisy channel by an error-correction
coding scheme.
The quantum analogs of Shannon entropy and Shannon noiseless coding
theorem are von Neumann entropy and Schumacher's noiseless channel
coding theorem, respectively. The von Neumann entropy
is defined to be
$ S(\brho) = - \operatorname{tr}(\brho \log\brho). $
Schumacher's noiseless channel coding theorem quantifies quantum
resources required to compress quantum states by von Neumann entropy
(\cite*{Sch95}).
Analogous to Shannon's noisy channel coding theorem, a theorem known as
Holevo--Schuma\-cher--Westmoreland theorem can be used to compute the
product quantum state capacity for some noisy channels
(\cite*{Hol98}; \cite*{SchWes97}). However, communications
over noisy quantum channels are much less understood than the classical
counterpart. It is an unsolved problem to determine quantum channel
capacity or the amount of quantum information that can be reliably
transmitted over noisy quantum channels. See \citet{Hay06} and
\citet{NieChu00}.

In spite of the above similarity, there are intrinsic differences
between classical information and quantum information.
Classical information can be distinguished and copied. For example, we
can identify different letters and produce an identical version of a
\mbox{digital} image for back-up.
However, quantum mechanics does not allow unknown quantum states to be
distinguished or copied exactly.
For example, we cannot reliably distinguish between quantum states
$|0\rangle$ and $(|0\rangle+ |1\rangle)/\sqrt{2}$. If we perform
measurement for quantum state $|0\rangle$,
the measurement result will be $0$ with probability 1, while measuring quantum
state $(|0\rangle+ |1\rangle)/\sqrt{2}$ yields measurements $0$ or
$1$ with equal probability.
A measurement result of $0$ cannot tell the identity of the quantum
state being measured. A theorem known as a no-cloning theorem states
that unknown quantum states cannot be copied exactly
(\cite*{WooZur82}; \cite*{NieChu00}).

As we discussed in Section~\ref{sec3.3}, quantum entanglement plays a crucial
role in strange quantum effects such as quantum teleportation,
violation of Bell's inequality, and superdense coding (\cite*
{Hay06}; \cite*{NieChu00}).
Entanglement is a new type of resource that differs vastly from the
traditional resources in classical information theory.
We are far from having a general theory to understand quantum
entanglement but encouraging progress made so far reveals the amazing
property and intriguing structure of entangled states and remarkable
connections between noisy quantum channels and entanglement transformation.
Consider\break 
quantum error-correction for reliable quantum computation and quantum
information processing.
Quantum error-correction is employed in quantum computation and quantum
communication to protect quantum information from loss due to quantum
noise and other errors like faulty quantum gates. 
Classical information uses redundancy to achieve error-correction, but
the no-cloning theorem presents an obstacle to copying quantum
information and formulating a theory of quantum error-correction based
on simple redundancy. Again quantum entanglement comes to the rescue.
It is forbidden to copy qubits but we can spread the information of one
qubit onto a highly entangled state of several qubits. \citet
{Sho95} first discovered the method of formulating a quantum
error-correction code by storing the information of one qubit onto a
highly entangled state of nine qubits. Over time several quantum
error-correction codes are proposed
(\cite*{CalSho96}; \cite*{Coretal98}; Steane, \citeyear{Ste96N1,Ste96N2}). These
quantum error-correction codes can protect quantum information against
quantum noise, 
and thus quantum noise likely poses no fundamental barrier to the
performance of large-scale quantum computing and quantum information processing.

Here is how quantum error-correction codes work. We consider the single
qubit case. First
assume that a qubit $\alpha_0 |0\rangle+ \alpha_1 |1\rangle$
is passed through a bit flip channel which flips the state of a qubit
from $|0\rangle$ to $|1\rangle$ and
from $|1\rangle$ to $|0\rangle$, each with probability $p$, and
leaves each of states $|0\rangle$ and $|1\rangle$
untouched with probability $1-p$. We describe a bit flip code that
protects the qubit against 
quantum noise from the bit flip channel.

We encode states $|0\rangle$ and $|1\rangle$ in three qubits, with
$|0\rangle$ encoded as $|000\rangle$ and $|1\rangle$ as $|111\rangle$.
Thus the qubit state $\alpha_0 |0\rangle+ \alpha_1 |1\rangle$ is
encoded in three qubits as
$\alpha_0 |000\rangle+ \alpha_1 |111\rangle$. We pass each of the
three qubits through
an independent copy of the bit flip channel, and assume that at most
one qubit is flipped. The following
simple two-step error-correction procedure can be used to recover the
correct quantum state.\looseness=1

Step 1. Perform a measurement on a specially constructed observable and
call the measurement result an
error syndrome. The error syndrome can inform us what error, if any,
occurred on the quantum state.
The observable has eigenvalues\vadjust{\goodbreak} $0$, $1$, $2$ and $3$, with
corresponding projection operators,
\begin{eqnarray*}
\bQ_0 &=& |000\rangle\langle000| + |111\rangle\langle111|  \quad
\mbox{no error}, \\
\bQ_1 &=& |100\rangle\langle100| + |011\rangle\langle011|  \\
&&\hspace*{-14pt}\mbox{bit flip on the first qubit},  \\
\bQ_2 &=& |010\rangle\langle010| + |101\rangle\langle101|   \\
&&\hspace*{-14pt}
\mbox{bit flip on the second qubit}, \\
\bQ_3 &=& |001\rangle\langle001| + |110\rangle\langle110|   \\
&&\hspace*{-14pt}
\mbox{bit flip on the third qubit}.
\end{eqnarray*}
If one of three qubits has one or no bit flip, the error syndrome will
be one of $0$, $1$, $2$ and $3$,
with $0$ corresponding to no flip, and $1$, $2$ and $3$ to a bit flip
on the first, second and third qubit,
respectively. For example, if the first qubit is flipped, the corrupted
state is
$|\psi\rangle= \alpha_0 |100\rangle+ \alpha_1 |011\rangle$. Since
$\langle\psi| \bQ_1 |\psi\rangle=1$ and
$\langle\psi| \bQ_j |\psi\rangle=0$ for $j \neq1$, in this case
the error syndrome is $1$. Although performing measurements usually
causes change to the quantum state,
the speciality of the constructed observable is that syndrome measurement
does not perturb the quantum state: it is easy to check that the state
is $|\psi\rangle$ both before and after
the syndrome measurement. While the syndrome provides information about
what flip error has occurred, it does not
contain any information about the state being protected, that is, it
does not allow us to deduce anything about the
amplitudes $\alpha_0$ and $\alpha_1$. Such a special property is the
generic feature of syndrome measurement.

Step 2. The error type supplied by the error syndrome can inform us
what procedure to use to recover
the original state. For example, error syndrome $1$ indicates a bit
flip on the first qubit, and a flip
on the first qubit again will perfectly recover the original state
$\alpha_0 |000\rangle+ \alpha_1 |111\rangle$.
The error syndrome $0$ implies no error and doing nothing, and error
syndromes $1$, $2$ and $3$ correspond
to a bit flip again on the first, second, third qubit, respectively.
The procedure will recover the original
state with perfect accuracy, if there is at most one bit flip in the
encoded three qubits. The probability that
more than one bit flipped is $p^3 + 3 p^2 (1-p) = 3p^2 - 2p^3$, which
is much smaller than the error
probability $p$ of making no-correction for the typical bit flip channel.
Thus the encoding and decoding scheme makes the storage and
transmission of the qubit more reliable.\looseness=1


Next we consider a more interesting noisy quantum channel: a phase flip
channel which, with probability $p$,
changes a qubit state $\alpha_0 |0\rangle+ \alpha_1 |1\rangle$ to\vadjust{\goodbreak}
$\alpha_0 |0\rangle- \alpha_1 |1\rangle$, and
with probability $1-p$, leaves alone the qubit. The following scheme is
to turn the phase flip channel into
a bit flip channel. Let $|+\rangle=( |0\rangle+ |1\rangle)/\sqrt
{2}$ and
$|-\rangle=( |0\rangle- |1\rangle)/\sqrt{2}$ be a qubit basis. The
phase flip channel leaves alone
states $|+\rangle$ and $|-\rangle$ with probability $1-p$ and changes
$|+\rangle$ to $|-\rangle$ and vice versa with probability $p$.
In other words, the phase flip channel with respect to the basis
$|+\rangle$ and $|-\rangle$ acts just like a bit flip channel
with respect to the basis $|0\rangle$ and $|1\rangle$. Thus we encode
$|0\rangle$ as $|{+}+{+}\rangle$ and $|1\rangle$ as
$|{-}-{-}\rangle$ for protection against phase flip errors. The operations
for encoding, error-detection and recovery
are the same as for the bit flip channel but with respect to the
$|+\rangle$ and $|-\rangle$ basis instead of
the $|0\rangle$ and $|1\rangle$ basis.

Last we describe Shor error-correction code.
It is a combination of the three-qubit phase flip and bit flip codes.
First use the phase flip code to encode states
$|0\rangle$ and $|1\rangle$ in three qubits, with $|0\rangle$
encoded as $|{+}+{+}\rangle$ and $|1\rangle$ as $|{-}-{-}\rangle$;
next, use the three-qubit bit flip code to encode each of these qubits,
with $|+\rangle$ encoded as
$(|000\rangle+ |111\rangle)/\sqrt{2}$ and $|-\rangle$ encoded as
$(|000\rangle- |111\rangle)/\sqrt{2}$.
The resulted nine-qubit code has codeworks as follows:
\begin{eqnarray*}
|0\rangle&\rightarrow&\frac{|000\rangle+ |111\rangle}{\sqrt{2}}
\frac{|000\rangle+ |111\rangle}{\sqrt{2}}
 \frac{|000\rangle+ |111\rangle}{\sqrt{2}},
\\
|1\rangle&\rightarrow&\frac{|000\rangle- |111\rangle}{\sqrt{2}}
\frac{|000\rangle- |111\rangle}{\sqrt{2}}
 \frac{|000\rangle- |111\rangle}{\sqrt{2}}.
\end{eqnarray*}
With the mixture of both phase flip and bit flip codes, the Shor
error-correction code can protect
against bit flip errors, phase flip errors, as well as a combined bit
and phase flip errors on any single qubit.
In fact it has been shown that this simple quantum error-correction
code can protect against the effects of any
completely arbitrary errors on a~single qubit (\cite*{Sho95}).



\section{Concluding Remarks}\label{sec7}

Quantum information science gains enormous attention in computer
science, mathematics, physical sciences and engineering,
and several interdisciplinary subfields are developing under the
umbrella of quantum information.
This paper 
reviews quantum computation and quantum information from a~statistical
perspective. We introduce concepts like
qubits, quantum gates and quantum circuits in quantum computation and
discuss quantum entanglement, quantum parallelism and
quantum error-cor\-rection in quantum computation and quantum information.
We present major quantum\vadjust{\goodbreak} algorithms and show their advantages over the
available classical algorithms. We illustrate quantum simulation
procedure and Monte Carlo methods in quantum simulation.
As classical computation and simulation are ubiquitous nowadays in
statistics, we expect quantum computation and quantum
simulation will have a~paramount role to play in modern statistics. 
This paper exposes the topics to statisticians and encourages more
statisticians to work in the fields. There are many statistical issues in
theoretical research as well as experimental work in quantum
computation, quantum simulation and quantum information.
For example, as measurement data collected in quantum \mbox{experiments}
require more and more sophisticated statistical methods for better
estimation, simulation and understanding,
it is imperative to develop good quantum statistics methods and quantum
simulation procedures and
study interrelationship and mutual impact between quantum estimation
and quantum simulation.
Since quantum computation is intrinsically random, and quantum
simulation employs Monte Carlo techniques, as we point out in Section~\ref{sec4.3} and \citet{Wan11},
it is important to provide sound statistical methods for analyzing
quantum algorithms and quantum simulation in general
and study high-order approximations to exponentiate Hamiltonians and
the efficiency of the resulted quantum simulation procedures in
particular. On the other hand,
quantum computation and quantum simulation have great potential to
revolutionize computational statistics. Below are a few cases in point.

\begin{enumerate}
\item[1.]
The ``random numbers'' generated by classical computers are
pseudo-random numbers in the\break sense that they are produced
by deterministic procedures and can be exactly repeated and perfectly
predicted given the deterministic schemes and the initial seeds.
On the contrary, superposition states enable quantum computers to
produce genuine random numbers. For example, measuring
$(|0\rangle+ |1\rangle)/\sqrt{2}$ yields $0$ and $1$ with equal
probability. In general we generate $b$-bit binary random numbers
$x=x_1\cdots x_b$, $x_j=0, 1$ as follows. Apply $b$ Hadamard gates to
$b$ qubits of $|0 \cdots0 \rangle$ to obtain
\begin{eqnarray*}
&&\frac{|0 \rangle+ |1 \rangle}{\sqrt{2}} \cdots\frac{|0 \rangle+
|1 \rangle}{\sqrt{2}}
= \frac{1}{\sqrt{2^b}} \sum_x |x\rangle, \\&&\quad x=x_1\cdots x_b,
x_j=0,1,
\end{eqnarray*}
where the sum is over all possible $2^b$ values of~$x$, and then
measure the obtained qubits and yield $b$-bit binary random numbers
$x=x_1\cdots x_b$ with equal probability. Quantum theory guarantees that
such random numbers are genuinely random.
Thus quantum computers are able to generate genuine random numbers and
perform true Monte Carlo simulation.
It is exciting to design general quantum random number generator and
study quantum Monte Carlo simulation. Perhaps we may need to re-examine
Monte Carlo simulation studies conducted by classical
computers.

\item[2.] It is interesting to investigate the potential of quantum
computation and quantum simulation for  computational statistics.
We expect that quantum computers may be much faster than classical
computers for computing some statistical problems. Moreover, quantum
computers may be able to
carry out some computational statistical tasks that are prohibitive by
classical computers. Specific examples are as follows:
(a) We may use the basic ideas of Grover's search algorithm to develop
fast quantum algorithms for implementing some statistical procedures.
For example, finding the median of a~huge data set is to search for
a~numerical value that separates the top and bottom halves of the data, and
quantum algorithms can offer quadratical speedup for calculating median
and trimmed mean. 
(b) With genuine random number generator and faster mean evaluation,
quantum computers may offer significant advantages over classical
computers for Monte Carlo
integration. For example, Monte Carlo integration in high dimensions
may be exponentially or
quadratically faster on quantum computers than on classical computers.
(c) It might be possible for quantum computers to carry out some
prohibiting statistical computing tasks like the Bayesian computation
discussed in Section~\ref{sec3}.
Some preliminary research along these lines may be found in \citet
{NayWu99} and \citet{Hei03}.

\item[3.] As quantum computation and quantum simulation are ideal for
simulating interacting particle systems like the Ising model, it is
fascinating to explore the interplay between quantum simulation and Markov
chain Monte Carlo methodology and the quantum potential to speed up
Markov chain based algorithms. In fact, it has been shown that
quantum walk based algorithms can offer quadratical speedup for certain
Markov chain based algorithms (\cite*{Magetal11}; \cite*
{Ric07}; \cite*{Sze},
and \cite*{WocAbe08}).
\end{enumerate}
%


Finally we point out that quantum computers are wonderful but it is
difficult to build quantum\vadjust{\goodbreak} computers with present technology. 
To build a quantum computer the physical apparatus must satisfy
requirements that the quantum system realized qubits needs to be
well isolated in order to retain its quantum properties and at the same
time the quantum system has to be accessible so that the qubits
can be manipulated to perform computations and measure output results.
The two opposing requirements are determined by
the strength of coupling of the quantum system to the external
entities. The coupling causes quantum decoherence. Decoherence refers to
the loss of coherence between the components of a quantum system or
quantum superposition from the interaction of the quantum system with
its environment. It is very crucial but challenging to control a
quantum system of qubits and correct the effects of decoherence in
quantum computation
and quantum information. Quantum computing has witnessed great advances
in recent years, and 
quantum computers of a handful of qubits and basic quantum communication devices have been built
in research laboratories (see \cite*{Bar12}; \cite*{ClaWil08}; \cite*
{DiCetal09}; \cite*{Johetal}; \cite*{Leeetal11}; \cite*{Neuetal08}), but there are technological hurdles in the
development of a quantum computer of large capacity.
History shows that scientific innovations and technological surprises
are a never-ending saga.
It is anticipated that quantum computers with a few dozen of qubits
will be built in near future. As we have discussed in Section~\ref{sec3.1},
such a quantum computer has capacity of a classical supercomputer.
We are very optimistic that someday quantum computers will be available
for statisticians to crunch numbers. For the time being, instead of
waiting in the sidelines for that to happen, statisticians should get
into the field of play.
It is time for us to dive into this frontier research and work with
scientists and engineers to speed up the arrival of practical quantum computers.
As a last note, in 2011 a Canadian company called D-Wave has sold the claimed first commercial
quantum computer of 128 qubits to the Lockheed-Martin corporation, despite the D-Wave's
quantum system being criticized as a black box. Large scale quantum computers may be years away,
but quantum computing is already here as a scientific endeavor to provoke deep thoughts and
integrate profound questions in physics and computer science.

\section*{Acknowledgments}

Wang's research was supported in part by NSF Grant DMS-10-05635.
He thanks editor David Madigan and two anonymous referees
for helpful comments and suggestions
which led to significant improvements in both substance and the
presentation of the paper.

%


\begin{thebibliography}{84}

\bibitem[\protect\citeauthoryear{Abrams and Lloyd}{1997}]{AbrLlo97}
%
\begin{barticle}[auto:STB|2012/01/27|08:30:54]
\bauthor{\bsnm{Abrams},~\bfnm{D.~S.}\binits{D.~S.}} \AND
\bauthor{\bsnm{Lloyd},~\bfnm{S.}\binits{S.}}
(\byear{1997}).
\btitle{Simulation of many-body Fermi systems on a quantum computer}.
\bjournal{Phys. Rev. Lett.}
\bvolume{79}
\bpages{2586--2589}.
\bptok{imsref}%
\end{barticle}
%
\endbibitem

\bibitem[\protect\citeauthoryear{Aharonov and Ta-Shma}{2003}]{AhaTaS03}
%
\begin{binproceedings}[mr]
\bauthor{\bsnm{Aharonov},~\bfnm{Dorit}\binits{D.}} \AND
\bauthor{\bsnm{Ta-Shma},~\bfnm{Amnon}\binits{A.}}
(\byear{2003}).
\btitle{Adiabatic quantum state generation and statistical zero knowledge}.
In \bbooktitle{Proceedings of the {T}hirty-{F}ifth {A}nnual {ACM} {S}ymposium
on {T}heory of {C}omputing}
\bpages{20--29}.
\bpublisher{ACM}, \baddress{New York} \bnote{(electronic)}.
\bid{doi={10.1145/780542.780546}, mr={2121066}}
\bptok{imsref}%
\end{binproceedings}
%
\endbibitem


\bibitem[\protect\citeauthoryear{Artiles, Gill and Gu{\c{t}}{\u
{a}}}{2005}]{ArtGilGut05}
%
\begin{barticle}[mr]
\bauthor{\bsnm{Artiles},~\bfnm{L.~M.}\binits{L.~M.}},
\bauthor{\bsnm{Gill},~\bfnm{R.~D.}\binits{R.~D.}} \AND
\bauthor{\bsnm{Gu{\c{t}}{\u{a}}},~\bfnm{M.~I.}\binits{M.~I.}}
(\byear{2005}).
\btitle{An invitation to quantum tomography}.
\bjournal{J. R. Stat. Soc. Ser. B Stat. Methodol.}
\bvolume{67}
\bpages{109--134}.
\bid{doi={10.1111/j.1467-9868.2005.00491.x}, issn={1369-7412}, mr={2136642}}
\bptok{imsref}%
\end{barticle}
%
\endbibitem

\bibitem[\protect\citeauthoryear{Aspect, Grangier and
Roger}{1981}]{AspGraRog81}
%
\begin{barticle}[auto:STB|2012/01/27|08:30:54]
\bauthor{\bsnm{Aspect},~\bfnm{A.}\binits{A.}},
\bauthor{\bsnm{Grangier},~\bfnm{P.}\binits{P.}} \AND
\bauthor{\bsnm{Roger},~\bfnm{G.}\binits{G.}}
(\byear{1981}).
\btitle{Experimental tests of realistic local theories via Bell's theorem}.
\bjournal{Phys. Rev. Lett.}
\bvolume{47}
\bpages{460--463}.
\bptok{imsref}%
\end{barticle}
%
\endbibitem

\bibitem[\protect\citeauthoryear{Aspect, Grangier and
Roger}{1982a}]{AspGraRog82N1}
%
\begin{barticle}[auto:STB|2012/01/27|08:30:54]
\bauthor{\bsnm{Aspect},~\bfnm{A.}\binits{A.}},
\bauthor{\bsnm{Grangier},~\bfnm{P.}\binits{P.}} \AND
\bauthor{\bsnm{Roger},~\bfnm{G.}\binits{G.}}
(\byear{1982}a).
\btitle{Experimental realization of Einstein--Podolsky--Rosen--Bohm
Gedankenexperiment: A new violation of Bell's inequalities}.
\bjournal{Phys. Rev. Lett.}
\bvolume{49}
\bpages{91--94}.
\bptok{imsref}%
\end{barticle}
%
\endbibitem

\bibitem[\protect\citeauthoryear{Aspect, Grangier and
Roger}{1982b}]{AspGraRog82N2}
%
\begin{barticle}[auto:STB|2012/01/27|08:30:54]
\bauthor{\bsnm{Aspect},~\bfnm{A.}\binits{A.}},
\bauthor{\bsnm{Grangier},~\bfnm{P.}\binits{P.}} \AND
\bauthor{\bsnm{Roger},~\bfnm{G.}\binits{G.}}
(\byear{1982}b).
\btitle{Experimental test of Bell's inequalities using time-varying analyzers}.
\bjournal{Phys. Rev. Lett.}
\bvolume{49}
\bpages{1804--1807}.
\bptok{imsref}%
\end{barticle}
%
\endbibitem

\bibitem[\protect\citeauthoryear{Aspuru-Guzik, Dutoi and
Head-Gordon}{2005}]{AspDutHea}
%
\begin{barticle}[auto:STB|2012/01/27|08:30:54]
\bauthor{\bsnm{Aspuru-Guzik},~\bfnm{A.~D.}\binits{A.~D.}},
\bauthor{\bsnm{Dutoi},~\bfnm{P.~J.~Love}\binits{P.~J.~L.}} \AND
\bauthor{\bsnm{Head-Gordon}, \bfnm{M.}\binits{M.}}
(\byear{2005}).
\btitle{Simulated quantum computation of molecular energies}.
\bjournal{Science} \bvolume{309} \bpages{1704}.
\bptok{imsref}%
\end{barticle}
%
\endbibitem

\bibitem[\protect\citeauthoryear{Barndorff-Nielsen, Gill and
Jupp}{2003}]{BarGilJup03}
%
\begin{barticle}[mr]
\bauthor{\bsnm{Barndorff-Nielsen},~\bfnm{Ole~E.}\binits{O.~E.}},
\bauthor{\bsnm{Gill},~\bfnm{Richard~D.}\binits{R.~D.}} \AND
\bauthor{\bsnm{Jupp},~\bfnm{Peter~E.}\binits{P.~E.}}
(\byear{2003}).
\btitle{On quantum statistical inference}.
\bjournal{J. R. Stat. Soc. Ser.~B Stat. Methodol.}
\bvolume{65}
\bpages{775--816}.
\bid{doi={10.1111/1467-9868.00415}, issn={1369-7412}, mr={2017871}}
\bptnote{check related}%
\bptok{imsref}%
\end{barticle}
%
\endbibitem

\bibitem[\protect\citeauthoryear{Barz et al.}{2012}]{Bar12}
\begin{barticle}[mr]
\bauthor{\bsnm{Barz},~\bfnm{Stefanie}\binits{S.}},
  \bauthor{\bsnm{Kashefi},~\bfnm{Elham}\binits{E.}},
  \bauthor{\bsnm{Broadbent},~\bfnm{Anne}\binits{A.}},
  \bauthor{\bsnm{Fitzsimons},~\bfnm{Joseph~F.}\binits{J.F.}},
  \bauthor{\bsnm{Zeilinger},~\bfnm{Anton}\binits{A.}} \AND
  \bauthor{\bsnm{Walther},~\bfnm{Philip}\binits{P.}}
(\byear{2012}).
\btitle{Demonstration of blind quantum computing}.
\bjournal{Science}
\bvolume{335}
\bpages{303--308}.
\bid{doi={10.1126/science.1214707}, issn={0036-8075}, mr={2919052}}
\bptok{imsref}%
\end{barticle}
\endbibitem

\bibitem[\protect\citeauthoryear{Bell}{1964}]{Bel64}
%
\begin{barticle}[auto:STB|2012/01/27|08:30:54]
\bauthor{\bsnm{Bell},~\bfnm{J.}\binits{J.}}
(\byear{1964}).
\btitle{On the Einstein Podolsky Rosen paradox}.
\bjournal{Physics}
\bvolume{1}
\bpages{195--200}.
\bptok{imsref}%
\end{barticle}
%
\endbibitem

\bibitem[\protect\citeauthoryear{Bennett et al.}{2002}]{Benetal02}
%
\begin{barticle}[mr]
\bauthor{\bsnm{Bennett},~\bfnm{C.~H.}\binits{C.~H.}},
\bauthor{\bsnm{Cirac},~\bfnm{J.~I.}\binits{J.~I.}},
\bauthor{\bsnm{Leifer},~\bfnm{M.~S.}\binits{M.~S.}},
\bauthor{\bsnm{Leung},~\bfnm{D.~W.}\binits{D.~W.}},
\bauthor{\bsnm{Linden},~\bfnm{N.}\binits{N.}},
\bauthor{\bsnm{Popescu},~\bfnm{S.}\binits{S.}} \AND
\bauthor{\bsnm{Vidal},~\bfnm{G.}\binits{G.}}
(\byear{2002}).
\btitle{Optimal simulation of two-qubit {H}amiltonians using general local
operations}.
\bjournal{Phys. Rev. A (3)}
\bvolume{66}
\bpages{012305}.
\bid{doi={10.1103/PhysRevA.66.012305}, issn={1050-2947}, mr={1929513}}
\bptok{imsref}%
\end{barticle}
%
\endbibitem

\bibitem[\protect\citeauthoryear{Berry et al.}{2007}]{Beretal07}
%
\begin{barticle}[mr]
\bauthor{\bsnm{Berry},~\bfnm{Dominic~W.}\binits{D.~W.}},
\bauthor{\bsnm{Ahokas},~\bfnm{Graeme}\binits{G.}},
\bauthor{\bsnm{Cleve},~\bfnm{Richard}\binits{R.}} \AND
\bauthor{\bsnm{Sanders},~\bfnm{Barry~C.}\binits{B.~C.}}
(\byear{2007}).
\btitle{Efficient quantum algorithms for simulating sparse {H}amiltonians}.
\bjournal{Comm. Math. Phys.}
\bvolume{270}
\bpages{359--371}.
\bid{doi={10.1007/s00220-006-0150-x}, issn={0010-3616}, mr={2276450}}
\bptok{imsref}%
\end{barticle}
%
\endbibitem

\bibitem[\protect\citeauthoryear{Boghosian and Taylor}{1998}]{BogTay98}
%
\begin{barticle}[mr]
\bauthor{\bsnm{Boghosian},~\bfnm{Bruce~M.}\binits{B.~M.}} \AND
\bauthor{\bsnm{Taylor},~\bfnm{Washington}\binits{W.} \bsuffix{IV}}
(\byear{1998}).
\btitle{Simulating quantum mechanics on a quantum computer}.
\bjournal{Phys. D}
\bvolume{120}
\bpages{30--42}.
\bid{doi={10.1016/S0167-2789(98)00042-6}, issn={0167-2789}, mr={1679863}}
\bptok{imsref}%
\end{barticle}
%
\endbibitem

\bibitem[\protect\citeauthoryear{Bohm}{1951}]{Boh51}
%
\begin{bbook}[auto:STB|2012/01/27|08:30:54]
\bauthor{\bsnm{Bohm},~\bfnm{D.}\binits{D.}}
(\byear{1951}).
\btitle{Quantum Theory}.
\bpublisher{Prentice-Hall}, \baddress{Englewood Cliffs, NJ}.
\bptok{imsref}%
\end{bbook}
%
\endbibitem

\bibitem[\protect\citeauthoryear{Butucea, Gu{\c{t}}{\u{a}} and
Artiles}{2007}]{ButGutArt07}
%
\begin{barticle}[mr]
\bauthor{\bsnm{Butucea},~\bfnm{Cristina}\binits{C.}},
\bauthor{\bsnm{Gu{\c{t}}{\u{a}}},~\bfnm{M{\u{a}}d{\u
{a}}lin}\binits{M.}} \AND
\bauthor{\bsnm{Artiles},~\bfnm{Luis}\binits{L.}}
(\byear{2007}).
\btitle{Minimax and adaptive estimation of the {W}igner\vadjust{\goodbreak} function in quantum
homodyne tomography with noisy data}.
\bjournal{Ann. Statist.}
\bvolume{35}
\bpages{465--494}.
\bid{doi={10.1214/009053606000001488}, issn={0090-5364}, mr={2336856}}
\bptok{imsref}%
\end{barticle}
%
\endbibitem

\bibitem[\protect\citeauthoryear{Calderbank and Shor}{1996}]{CalSho96}
%
\begin{barticle}[auto:STB|2012/01/27|08:30:54]
\bauthor{\bsnm{Calderbank},~\bfnm{A.~R.}\binits{A.~R.}} \AND
\bauthor{\bsnm{Shor},~\bfnm{P.~W.}\binits{P.~W.}}
(\byear{1996}).
\btitle{Good quantum error-correcting codes exist}.
\bjournal{Phys. Rev. A}
\bvolume{54}
\bpages{1098--1105}.
\bptok{imsref}%
\end{barticle}
%
\endbibitem

\bibitem[\protect\citeauthoryear{Childs}{2010}]{Chi10}
%
\begin{barticle}[mr]
\bauthor{\bsnm{Childs},~\bfnm{Andrew~M.}\binits{A.~M.}}
(\byear{2010}).
\btitle{On the relationship between continuous- and discrete-time quantum
walk}.
\bjournal{Comm. Math. Phys.}
\bvolume{294}
\bpages{581--603}.
\bid{doi={10.1007/s00220-009-0930-1}, issn={0010-3616}, mr={2579466}}
\bptok{imsref}%
\end{barticle}
%
\endbibitem

\bibitem[\protect\citeauthoryear{Childs et al.}{2003}]{Chietal}
%
\begin{binproceedings}[auto:STB|2012/01/27|08:30:54]
\bauthor{\bsnm{Childs},~\bfnm{A.~M.}\binits{A.~M.}},
\bauthor{\bsnm{Cleve},~\bfnm{R.}\binits{R.}},
\bauthor{\bsnm{Deotto},~\bfnm{E.}\binits{E.}},
\bauthor{\bsnm{Farhi},~\bfnm{E.}\binits{E.}},
\bauthor{\bsnm{Gutmann},~\bfnm{S.}\binits{S.}} \AND
\bauthor{\bsnm{Spielman},~\bfnm{D.~A.}\binits{D.~A.}}
(\byear{2003}).
\btitle{Exponential algorithmic speedup by quantum walk}.
In
\bbooktitle{Proc. 35th ACM Symposium on Theory of Computing}
\bpages{59--68}.
\bpublisher{ACM Press}, \baddress{New York}.
\bptok{imsref}%
\end{binproceedings}
%
\endbibitem



\bibitem[\protect\citeauthoryear{Clarke and Wilhelm}{2008}]{ClaWil08}
%
\begin{barticle}[pbm]
\bauthor{\bsnm{Clarke},~\bfnm{John}\binits{J.}} \AND
\bauthor{\bsnm{Wilhelm},~\bfnm{Frank~K.}\binits{F.~K.}}
(\byear{2008}).
\btitle{Superconducting quantum bits}.
\bjournal{Nature}
\bvolume{453}
\bpages{1031--1042}.
\bid{doi={10.1038/nature07128}, issn={1476-4687}, pii={nature07128},
pmid={18563154}}
\bptok{imsref}%
\end{barticle}
%
\endbibitem

\bibitem[\protect\citeauthoryear{Clauser et al.}{1969}]{Claetal69}
%
\begin{barticle}[auto:STB|2012/01/27|08:30:54]
\bauthor{\bsnm{Clauser},~\bfnm{J.~F.}\binits{J.~F.}},
\bauthor{\bsnm{Horne},~\bfnm{M.~A.}\binits{M.~A.}},
\bauthor{\bsnm{Shimony},~\bfnm{A.}\binits{A.}} \AND
\bauthor{\bsnm{Holt},~\bfnm{R.~A.}\binits{R.~A.}}
(\byear{1969}).
\btitle{Proposed experiment to test local hidden-variable theories}.
\bjournal{Phys. Rev. Lett.}
\bvolume{23}
\bpages{880--884}.
\bptok{imsref}%
\end{barticle}
%
\endbibitem

\bibitem[\protect\citeauthoryear{Cory et al.}{1998}]{Coretal98}
%
\begin{barticle}[auto:STB|2012/01/27|08:30:54]
\bauthor{\bsnm{Cory},~\bfnm{D.~G.}\binits{D.~G.}},
\bauthor{\bsnm{Mass},~\bfnm{W.}\binits{W.}},
\bauthor{\bsnm{Price},~\bfnm{M.}\binits{M.}},
\bauthor{\bsnm{Knill},~\bfnm{E.}\binits{E.}},
\bauthor{\bsnm{Laflamme},~\bfnm{R.}\binits{R.}},
\bauthor{\bsnm{Zurek},~\bfnm{W.~H.}\binits{W.~H.}},
\bauthor{\bsnm{Havel},~\bfnm{T.~F.}\binits{T.~F.}} \AND
\bauthor{\bsnm{Somaroo},~\bfnm{S.~S.}\binits{S.~S.}}
(\byear{1998}).
\btitle{Experimental quantum error correction}.
\bjournal{Phys. Rev. Lett.}
\bvolume{81}
\bpages{2152--2155}.
\bptok{imsref}%
\end{barticle}
%
\endbibitem

\bibitem[\protect\citeauthoryear{Crandall and Pomerance}{2001}]{CraPom01}
%
\begin{bbook}[mr]
\bauthor{\bsnm{Crandall},~\bfnm{Richard}\binits{R.}} \AND
\bauthor{\bsnm{Pomerance},~\bfnm{Carl}\binits{C.}}
(\byear{2001}).
\btitle{Prime Numbers: A~Computational Perspective}.
\bpublisher{Springer}, \baddress{New York}.
\bid{mr={1821158}}
\bptok{imsref}%
\end{bbook}
%
\endbibitem

\bibitem[\protect\citeauthoryear{Deutsch}{1985}]{Deu85}
%
\begin{barticle}[mr]
\bauthor{\bsnm{Deutsch},~\bfnm{D.}\binits{D.}}
(\byear{1985}).
\btitle{Quantum theory, the {C}hurch--{T}uring principle and the universal
quantum computer}.
\bjournal{Proc. R. Soc. Lond. Ser. A}
\bvolume{400}
\bpages{97--117}.
\bid{issn={0962-8444}, mr={0801665}}
\bptok{imsref}%
\end{barticle}
%
\endbibitem

\bibitem[\protect\citeauthoryear{DiCarlo et al.}{2009}]{DiCetal09}
%
\begin{barticle}[pbm]
\bauthor{\bsnm{DiCarlo},~\bfnm{L.}\binits{L.}},
\bauthor{\bsnm{Chow},~\bfnm{J.~M.}\binits{J.~M.}},
\bauthor{\bsnm{Gambetta},~\bfnm{J.~M.}\binits{J.~M.}},
\bauthor{\bsnm{Bishop},~\bfnm{Lev~S.}\binits{L.~S.}},
\bauthor{\bsnm{Johnson},~\bfnm{B.~R.}\binits{B.~R.}},
\bauthor{\bsnm{Schuster},~\bfnm{D.~I.}\binits{D.~I.}},
\bauthor{\bsnm{Majer},~\bfnm{J.}\binits{J.}},
\bauthor{\bsnm{Blais},~\bfnm{A.}\binits{A.}},
\bauthor{\bsnm{Frunzio},~\bfnm{L.}\binits{L.}},
\bauthor{\bsnm{Girvin},~\bfnm{S.~M.}\binits{S.~M.}} \AND
\bauthor{\bsnm{Schoelkopf},~\bfnm{R.~J.}\binits{R.~J.}}
(\byear{2009}).
\btitle{Demonstration of two-qubit algorithms with a superconducting quantum
processor}.
\bjournal{Nature}
\bvolume{460}
\bpages{240--244}.
\bid{doi={10.1038/nature08121}, issn={1476-4687}, pii={nature08121},
pmid={19561592}}
\bptok{imsref}%
\end{barticle}
%
\endbibitem

\bibitem[\protect\citeauthoryear{DiVincenzo}{1995}]{DiV95}
%
\begin{barticle}[mr]
\bauthor{\bsnm{DiVincenzo},~\bfnm{David~P.}\binits{D.~P.}}
(\byear{1995}).
\btitle{Quantum computation}.
\bjournal{Science}
\bvolume{270}
\bpages{255--261}.
\bid{issn={0036-8075}, mr={1355956}}
\bptok{imsref}%
\end{barticle}
%
\endbibitem


\bibitem[\protect\citeauthoryear{Einstein, Podolsky and
Rosen}{1935}]{EinPodRos35}
%
\begin{barticle}[auto:STB|2012/01/27|08:30:54]
\bauthor{\bsnm{Einstein},~\bfnm{A.}\binits{A.}},
\bauthor{\bsnm{Podolsky},~\bfnm{B.}\binits{B.}} \AND
\bauthor{\bsnm{Rosen},~\bfnm{N.}\binits{N.}}
(\byear{1935}).
\btitle{Can quantum-mechanical description of physical reality be considered
complete?}
\bjournal{Phys. Rev.}
\bvolume{47}
\bpages{777--780}.
\bptok{imsref}%
\end{barticle}
%
\endbibitem

\bibitem[\protect\citeauthoryear{Feynman}{1981/82}]{Fey81}
%
\begin{barticle}[mr]
\bauthor{\bsnm{Feynman},~\bfnm{Richard~P.}\binits{R.~P.}}
(\byear{1981/82}).
\btitle{Simulating physics with computers}.
\bjournal{Internat. J. Theoret. Phys.}
\bvolume{21}
\bpages{467--488}.
\bid{doi={10.1007/BF02650179}, issn={0020-7748}, mr={0658311}}
\bptnote{check year}%
\bptok{imsref}%
\end{barticle}
%
\endbibitem

\bibitem[\protect\citeauthoryear{Freedman, Kitaev and
Wang}{2002}]{FreKitWan02}
%
\begin{barticle}[mr]
\bauthor{\bsnm{Freedman},~\bfnm{Michael~H.}\binits{M.~H.}},
\bauthor{\bsnm{Kitaev},~\bfnm{Alexei}\binits{A.}} \AND
\bauthor{\bsnm{Wang},~\bfnm{Zhenghan}\binits{Z.}}
(\byear{2002}).
\btitle{Simulation of topological field theories by quantum computers}.
\bjournal{Comm. Math. Phys.}
\bvolume{227}
\bpages{587--603}.
\bid{doi={10.1007/s002200200635}, issn={0010-3616}, mr={1910832}}
\bptok{imsref}%
\end{barticle}
%
\endbibitem

\bibitem[\protect\citeauthoryear{Griffiths}{2004}]{Gri}
%
\begin{bmisc}[auto:STB|2012/01/27|08:30:54]
\bauthor{\bsnm{Griffiths},~\bfnm{D.~J.}\binits{D.~J.}}
(\byear{2004}).
\bhowpublished{\textit{Introduction to Quantum Mechanics}, 2nd ed. Benjamin Cummings, San Francisco, CA}.
\bptok{imsref}%
\end{bmisc}
%
\endbibitem

\bibitem[\protect\citeauthoryear{Grover}{1996}]{Gro96}
%
\begin{binproceedings}[mr]
\bauthor{\bsnm{Grover},~\bfnm{Lov~K.}\binits{L.~K.}}
(\byear{1996}).
\btitle{A fast quantum mechanical algorithm for database search}.
In \bbooktitle{Proceedings of the {T}wenty-Eighth {A}nnual {ACM}
{S}ymposium on
the {T}heory of {C}omputing ({P}hiladelphia, {PA}, 1996)}
\bpages{212--219}.
\bpublisher{ACM}, \baddress{New York}.
\bid{doi={10.1145/237814.237866}, mr={1427516}}
\bptok{imsref}%
\end{binproceedings}
%
\endbibitem

\bibitem[\protect\citeauthoryear{Grover}{1997}]{Gro97}
%
\begin{barticle}[auto:STB|2012/01/27|08:30:54]
\bauthor{\bsnm{Grover},~\bfnm{L.~K.}\binits{L.~K.}}
(\byear{1997}).
\btitle{Quantum mechanics helps in searching for a needle in a haystack}.
\bjournal{Phys. Rev. Lett.}
\bvolume{79}
\bpages{325--328}.
\bptok{imsref}%
\end{barticle}
%
\endbibitem

\bibitem[\protect\citeauthoryear{Hayashi}{2006}]{Hay06}
%
\begin{bbook}[mr]
\bauthor{\bsnm{Hayashi},~\bfnm{Masahito}\binits{M.}}
(\byear{2006}).
\btitle{Quantum Information: An Introduction}.
\bpublisher{Springer}, \baddress{Berlin}.
\bnote{Translated from the 2003 Japanese original}.
\bid{mr={2228302}}
\bptok{imsref}%
\end{bbook}
%
\endbibitem

\bibitem[\protect\citeauthoryear{Heinrich}{2003}]{Hei03}
%
\begin{barticle}[mr]
\bauthor{\bsnm{Heinrich},~\bfnm{Stefan}\binits{S.}}
(\byear{2003}).
\btitle{From {M}onte {C}arlo to quantum computation}.
\bjournal{Math. Comput. Simulation}
\bvolume{62}
\bpages{219--230}.
\bid{doi={10.1016/S0378-4754(02)00239-2}, issn={0378-4754}, mr={1988372}}
\bptok{imsref}%
\end{barticle}
%
\endbibitem

\bibitem[\protect\citeauthoryear{Holevo}{1982}]{Hol82}
%
\begin{bbook}[mr]
\bauthor{\bsnm{Holevo},~\bfnm{A.~S.}\binits{A.~S.}}
(\byear{1982}).
\btitle{Probabilistic and Statistical Aspects of Quantum Theory}.
\bseries{North-Holland Series in Statistics and Probability}
\bvolume{1}.
\bpublisher{North-Holland}, \baddress{Amsterdam}.
\bnote{Translated from the Russian by the author}.
\bid{mr={0681693}}
\bptok{imsref}%
\end{bbook}
%
\endbibitem

\bibitem[\protect\citeauthoryear{Holevo}{1998}]{Hol98}
%
\begin{barticle}[mr]
\bauthor{\bsnm{Holevo},~\bfnm{A.~S.}\binits{A.~S.}}
(\byear{1998}).
\btitle{The capacity of the quantum channel with general signal states}.
\bjournal{IEEE Trans. Inform. Theory}
\bvolume{44}
\bpages{269--273}.
\bid{doi={10.1109/18.651037}, issn={0018-9448}, mr={1486663}}
\bptok{imsref}%
\end{barticle}
%
\endbibitem

\bibitem[\protect\citeauthoryear{Huber and Ronchetti}{2009}]{HubRon09}
%
\begin{bbook}[mr]
\bauthor{\bsnm{Huber},~\bfnm{Peter~J.}\binits{P.~J.}} \AND
\bauthor{\bsnm{Ronchetti},~\bfnm{Elvezio~M.}\binits{E.~M.}}
(\byear{2009}).
\btitle{Robust Statistics},
\bedition{2nd} ed.
\bpublisher{Wiley}, \baddress{Hoboken, NJ}.
\bid{doi={10.1002/9780470434697}, mr={2488795}}
\bptnote{check related}%
\bptok{imsref}%
\end{bbook}
%
\endbibitem

\bibitem[\protect\citeauthoryear{Jan{\'e} et al.}{2003}]{Janetal03}
%
\begin{barticle}[mr]
\bauthor{\bsnm{Jan{\'e}},~\bfnm{E.}\binits{E.}},
\bauthor{\bsnm{Vidal},~\bfnm{G.}\binits{G.}},
\bauthor{\bsnm{D{\"u}r},~\bfnm{W.}\binits{W.}},
\bauthor{\bsnm{Zoller},~\bfnm{P.}\binits{P.}} \AND
\bauthor{\bsnm{Cirac},~\bfnm{J.~I.}\binits{J.~I.}}
(\byear{2003}).
\btitle{Simulation of quantum dynamics with quantum optical systems}.
\bjournal{Quantum Inf. Comput.}
\bvolume{3}
\bpages{15--37}.
\bid{issn={1533-7146}, mr={1965173}}
\bptok{imsref}%
\end{barticle}
%
\endbibitem

\bibitem[\protect\citeauthoryear{Johnson et al.}{2011}]{Johetal}
%
\begin{barticle}[auto:STB|2012/01/27|08:30:54]
\bauthor{\bsnm{Johnson},~\bfnm{M.~W.}\binits{M.~W.}},
\bauthor{\bsnm{Amin},~\bfnm{M.~H.~S.}\binits{M.~H.~S.}},
\bauthor{\bsnm{Gildert},~\bfnm{S.}\binits{S.}},
\bauthor{\bsnm{Lanting},~\bfnm{T.}\binits{T.}},
\bauthor{\bsnm{Hamze},~\bfnm{F.}\binits{F.}},
\bauthor{\bsnm{Dickson},~\bfnm{N.}\binits{N.}},
\bauthor{\bsnm{Harris},~\bfnm{R.}\binits{R.}},
\bauthor{\bsnm{Berkley},~\bfnm{A.~J.}\binits{A.~J.}},
\bauthor{\bsnm{Johansson},~\bfnm{J.}\binits{J.}},
\bauthor{\bsnm{Bunyk},~\bfnm{P.}\binits{P.}},
\bauthor{\bsnm{Chapple},~\bfnm{E.~M.}\binits{E.~M.}},
\bauthor{\bsnm{Enderud},~\bfnm{C.}\binits{C.}},
\bauthor{\bsnm{Hilton},~\bfnm{J.~P.}\binits{J.~P.}},
\bauthor{\bsnm{Karimi},~\bfnm{K.}\binits{K.}},
\bauthor{\bsnm{Ladizinsky},~\bfnm{E.}\binits{E.}},
\bauthor{\bsnm{Ladizinsky},~\bfnm{N.}\binits{N.}},
\bauthor{\bsnm{Oh},~\bfnm{T.}\binits{T.}},
\bauthor{\bsnm{Perminov},~\bfnm{I.}\binits{I.}},
\bauthor{\bsnm{Rich},~\bfnm{C.}\binits{C.}},
\bauthor{\bsnm{Thom},~\bfnm{M. C.}\binits{M. C.}},
\bauthor{\bsnm{Tolkacheva},~\bfnm{E.}\binits{E.}},
\bauthor{\bsnm{Truncik},~\bfnm{C. J. S.}\binits{C. J. S.}},
\bauthor{\bsnm{Uchaikin},~\bfnm{S.}\binits{S.}},
\bauthor{\bsnm{Wang},~\bfnm{J.}\binits{J.}},
\bauthor{\bsnm{Wilson},~\bfnm{B.}\binits{B.}},
\bauthor{\bsnm{Rose},~\bfnm{G.}\binits{G.}} \betal{et~al.}
(\byear{2011}).
\btitle{Quantum annealing with manufactured spins}. \bjournal{Nature}
\bvolume{473}
\bpages{194--198}.
\bptok{imsref}%
\end{barticle}
%
\endbibitem

\bibitem[\protect\citeauthoryear{Kato}{1978}]{Kat78}
%
\begin{bincollection}[mr]
\bauthor{\bsnm{Kato},~\bfnm{Tosio}\binits{T.}}
(\byear{1978}).
\btitle{Trotter's product formula for an arbitrary pair of self-adjoint
contraction semigroups}.
In \bbooktitle{Topics in Functional Analysis (Essays Dedicated to {M}. {G}.
{K}re\u\i N on the Occasion of His 70th Birthday)}.
\bseries{Adv. in Math. Suppl. Stud.}
\bvolume{3}
\bpages{185--195}.
\bpublisher{Academic Press}, \baddress{New York}.
\bid{mr={0538020}}
\bptok{imsref}%
\end{bincollection}
%
\endbibitem


\bibitem[\protect\citeauthoryear{Kiefer}{2004}]{Kie04}
%
\begin{bincollection}[mr]
\bauthor{\bsnm{Kiefer},~\bfnm{Claus}\binits{C.}}
(\byear{2004}).
\btitle{On the interpretation of quantum theory---from {C}openhagen to the
present day}.
In \bbooktitle{Time, Quantum and Information}
\bpages{291--299}.
\bpublisher{Springer}, \baddress{Berlin}.
\bid{mr={2180154}}
\bptnote{check year}%
\bptok{imsref}%
\end{bincollection}
%
\endbibitem

\bibitem[\protect\citeauthoryear{Lee et al.}{2011}]{Leeetal11}
%
\begin{barticle}[pbm]
\bauthor{\bsnm{Lee},~\bfnm{Noriyuki}\binits{N.}},
\bauthor{\bsnm{Benichi},~\bfnm{Hugo}\binits{H.}},
\bauthor{\bsnm{Takeno},~\bfnm{Yuishi}\binits{Y.}},
\bauthor{\bsnm{Takeda},~\bfnm{Shuntaro}\binits{S.}},
\bauthor{\bsnm{Webb},~\bfnm{James}\binits{J.}},
\bauthor{\bsnm{Huntington},~\bfnm{Elanor}\binits{E.}} \AND
\bauthor{\bsnm{Furusawa},~\bfnm{Akira}\binits{A.}}
(\byear{2011}).
\btitle{Teleportation of nonclassical wave packets of light}.
\bjournal{Science}
\bvolume{332}
\bpages{330--\break333}.
\bid{doi={10.1126/science.1201034}, issn={1095-9203}, pii={332/6027/330},
pmid={21493853}}
\bptok{imsref}%
\end{barticle}
%
\endbibitem

\bibitem[\protect\citeauthoryear{Lloyd}{1996}]{Llo96}
%
\begin{barticle}[mr]
\bauthor{\bsnm{Lloyd},~\bfnm{Seth}\binits{S.}}
(\byear{1996}).
\btitle{Universal quantum simulators}.
\bjournal{Science}
\bvolume{273}
\bpages{1073--1078}.
\bid{issn={0036-8075}, mr={1407944}}
\bptok{imsref}%
\end{barticle}
%
\endbibitem

\bibitem[\protect\citeauthoryear{Magniez et al.}{2011}]{Magetal11}
%
\begin{barticle}[mr]
\bauthor{\bsnm{Magniez},~\bfnm{Fr{\'e}d{\'e}ric}\binits{F.}},
\bauthor{\bsnm{Nayak},~\bfnm{Ashwin}\binits{A.}},
\bauthor{\bsnm{Roland},~\bfnm{J{\'e}r{\'e}mie}\binits{J.}} \AND
\bauthor{\bsnm{Santha},~\bfnm{Miklos}\binits{M.}}
(\byear{2011}).
\btitle{Search via quantum walk}.
\bjournal{SIAM J. Comput.}
\bvolume{40}
\bpages{142--164}.
\bid{doi={10.1137/090745854}, issn={0097-5397}, mr={2783206}}
\bptok{imsref}%
\end{barticle}
%
\endbibitem

\bibitem[\protect\citeauthoryear{Mariantoni et al.}{2011}]{Maretal}
%
\begin{bmisc}[auto:STB|2012/01/27|08:30:54]
\bauthor{\bsnm{Mariantoni},~\bfnm{M.}\binits{M.}},
\bauthor{\bsnm{Wang},~\bfnm{H.}\binits{H.}},
\bauthor{\bsnm{Yamamoto},~\bfnm{T.}\binits{T.}},
\bauthor{\bsnm{Neeley},~\bfnm{M.}\binits{M.}},
\bauthor{\bsnm{Bialczak1},~\bfnm{R.~C.}\binits{R.~C.}},
\bauthor{\bsnm{Chen},~\bfnm{Y.}\binits{Y.}},
\bauthor{\bsnm{Lenander},~\bfnm{M.}\binits{M.}},
\bauthor{\bsnm{Lucero},~\bfnm{E.}\binits{E.}},
\bauthor{\bsnm{O'Connell},~\bfnm{A.~D.}\binits{A.~D.}},
\bauthor{\bsnm{Sank},~\bfnm{D.}\binits{D.}},
\bauthor{\bsnm{Weides},~\bfnm{M.}\binits{M.}},
\bauthor{\bsnm{Wenner},~\bfnm{J.}\binits{J.}},
\bauthor{\bsnm{Yin},~\bfnm{Y.}\binits{Y.}},
\bauthor{\bsnm{Zhao},~\bfnm{J.}\binits{J.}},
\bauthor{\bsnm{Korotkov},~\bfnm{A.~N.}\binits{A.~N.}},
\bauthor{\bsnm{Cleland},~\bfnm{A.~N.}\binits{A.~N.}} \AND
\bauthor{\bsnm{Martinis},~\bfnm{J.~M.}\binits{J.~M.}}
(\byear{2011}).
\bhowpublished{Implementing the quantum von Neumann architecture with
superconducting circuits. \textit{Science} \textbf{7} 1208517}.
\bptok{imsref}%
\end{bmisc}
%
\endbibitem

\bibitem[\protect\citeauthoryear{Menezes, van Oorschot and
Vanstone}{1996}]{MenvanVan96}
%
\begin{bbook}[auto:STB|2012/01/27|08:30:54]
\bauthor{\bsnm{Menezes},~\bfnm{A.}\binits{A.}}, \bauthor
{\bparticle{van}
\bsnm{Oorschot},~\bfnm{P.~C.}\binits{P.~C.}} \AND
\bauthor{\bsnm{Vanstone},~\bfnm{S.~A.}\binits{S.~A.}}
(\byear{1996}).
\btitle{Handbook of Applied Cryptography}.
\bpublisher{CRC Press}, \baddress{New York}.
\bptok{imsref}%
\end{bbook}
%
\endbibitem

\bibitem[\protect\citeauthoryear{Nayak and Wu}{1999}]{NayWu99}
%
\begin{bincollection}[mr]
\bauthor{\bsnm{Nayak},~\bfnm{Ashwin}\binits{A.}} \AND
\bauthor{\bsnm{Wu},~\bfnm{Felix}\binits{F.}}
(\byear{1999}).
\btitle{The quantum query complexity of approximating the median and related
statistics}.
In \bbooktitle{Annual {ACM} {S}ymposium on {T}heory of {C}omputing ({A}tlanta,
{GA}, 1999)}
\bpages{384--393}.
\bpublisher{ACM}, \baddress{New York}
\bnote{(electronic)}.
\bid{doi={10.1145/301250.301349}, mr={1798059}}
\bptok{imsref}%
\end{bincollection}
%
\endbibitem

\bibitem[\protect\citeauthoryear{Neumann et al.}{2008}]{Neuetal08}
%
\begin{barticle}[pbm]
\bauthor{\bsnm{Neumann},~\bfnm{P.}\binits{P.}},
\bauthor{\bsnm{Mizuochi},~\bfnm{N.}\binits{N.}},
\bauthor{\bsnm{Rempp},~\bfnm{F.}\binits{F.}},
\bauthor{\bsnm{Hemmer},~\bfnm{P.}\binits{P.}},
\bauthor{\bsnm{Watanabe},~\bfnm{H.}\binits{H.}},
\bauthor{\bsnm{Yamasaki},~\bfnm{S.}\binits{S.}},
\bauthor{\bsnm{Jacques},~\bfnm{V.}\binits{V.}},
\bauthor{\bsnm{Gaebel},~\bfnm{T.}\binits{T.}},
\bauthor{\bsnm{Jelezko},~\bfnm{F.}\binits{F.}} \AND
\bauthor{\bsnm{Wrachtrup},~\bfnm{J.}\binits{J.}}
(\byear{2008}).
\btitle{Multipartite entanglement among single spins in diamond}.
\bjournal{Science}
\bvolume{320}
\bpages{1326--1329}.
\bid{doi={10.1126/science.1157233}, issn={1095-9203}, pii={320/5881/1326},
pmid={18535240}}
\bptok{imsref}%
\end{barticle}
%
\endbibitem

\bibitem[\protect\citeauthoryear{Nielsen and Chuang}{2000}]{NieChu00}
%
\begin{bbook}[mr]
\bauthor{\bsnm{Nielsen},~\bfnm{Michael~A.}\binits{M.~A.}} \AND
\bauthor{\bsnm{Chuang},~\bfnm{Isaac~L.}\binits{I.~L.}}
(\byear{2000}).
\btitle{Quantum Computation and Quantum Information}.
\bpublisher{Cambridge Univ. Press}, \baddress{Cambridge}.
\bid{mr={1796805}}
\bptok{imsref}%
\end{bbook}
%
\endbibitem


\bibitem[\protect\citeauthoryear{Nightingale and Umrigar}{1999}]{autokey52}
%
\begin{bbook}[mr]
\beditor{\bsnm{Nightingale},~\bfnm{M.~P.}\binits{M.~P.}} \AND
\beditor{\bsnm{Umrigar},~\bfnm{C.~J.}\binits{C.~J.}}, eds.
(\byear{1999}).
\btitle{Quantum {M}onte {C}arlo Methods in Physics and Chemistry}.
\bseries{NATO Science Series C: Mathematical and Physical Sciences}
\bvolume{525}.
\bpublisher{Kluwer Academic}, \baddress{Dordrecht}.
\bid{mr={1712250}}
\bptok{imsref}%
\end{bbook}
%
\endbibitem

\bibitem[\protect\citeauthoryear{Nussbaum and Szko{\l}a}{2009}]{NusSzk09}
%
\begin{barticle}[mr]
\bauthor{\bsnm{Nussbaum},~\bfnm{Michael}\binits{M.}} \AND
\bauthor{\bsnm{Szko{\l}a},~\bfnm{Arleta}\binits{A.}}
(\byear{2009}).
\btitle{The {C}hernoff lower bound for symmetric quantum hypothesis testing}.
\bjournal{Ann. Statist.}
\bvolume{37}
\bpages{1040--1057}.
\bid{doi={10.1214/08-AOS593}, issn={0090-5364}, mr={2502660}}
\bptok{imsref}%
\end{barticle}
%
\endbibitem

\bibitem[\protect\citeauthoryear{Parthasarathy}{1992}]{Par92}
%
\begin{bbook}[mr]
\bauthor{\bsnm{Parthasarathy},~\bfnm{K.~R.}\binits{K.~R.}}
(\byear{1992}).
\btitle{An Introduction to Quantum Stochastic Calculus}.
\bseries{Monographs in Mathematics}
\bvolume{85}.
\bpublisher{Birkh\"auser}, \baddress{Basel}.
\bid{mr={1164866}}
\bptok{imsref}%
\end{bbook}
%
\endbibitem

\bibitem[\protect\citeauthoryear{Richter}{2007}]{Ric07}
%
\begin{barticle}[auto:STB|2012/01/27|08:30:54]
\bauthor{\bsnm{Richter},~\bfnm{P.~C.}\binits{P.~C.}}
(\byear{2007}).
\btitle{Quantum speedup of classical mixing processes}.
\bjournal{Phys. Rev. A}
\bvolume{76}
\bpages{042306}.
\bptok{imsref}%
\end{barticle}
%
\endbibitem

\bibitem[\protect\citeauthoryear{Rivest, Shamir and
Adleman}{1978}]{RivShaAdl78}
%
\begin{barticle}[mr]
\bauthor{\bsnm{Rivest},~\bfnm{R.~L.}\binits{R.~L.}},
\bauthor{\bsnm{Shamir},~\bfnm{A.}\binits{A.}} \AND
\bauthor{\bsnm{Adleman},~\bfnm{L.}\binits{L.}}
(\byear{1978}).
\btitle{A~method for obtaining digital signatures and public-key
cryptosystems}.
\bjournal{Comm. ACM}
\bvolume{21}
\bpages{120--126}.
\bid{doi={10.1145/359340.359342}, issn={0001-0782}, mr={0700103}}
\bptok{imsref}%
\end{barticle}
%
\endbibitem

\bibitem[\protect\citeauthoryear{Rousseau}{2008}]{Rou08}
%
\begin{barticle}[mr]
\bauthor{\bsnm{Rousseau},~\bfnm{V.~G.}\binits{V.~G.}}
(\byear{2008}).
\btitle{Stochastic {G}reen function algorithm}.
\bjournal{Phys. Rev. E (3)}
\bvolume{77}
\bpages{056705}.
\bid{doi={10.1103/PhysRevE.77.056705}, issn={1539-3755}, mr={2495497}}
\bptok{imsref}%
\end{barticle}
%
\endbibitem

\bibitem[\protect\citeauthoryear{Sakurai and Napolitano}{2010}]{SakNap10}
%
\begin{bbook}[auto:STB|2012/01/27|08:30:54]
\bauthor{\bsnm{Sakurai},~\bfnm{J.~J.}\binits{J.~J.}} \AND
\bauthor{\bsnm{Napolitano},~\bfnm{J.}\binits{J.}}
(\byear{2010}).
\btitle{Modern Quantum Mechanics}, \bedition{2}nd ed.
\bpublisher{Addison-Wesley}, \baddress{Reading, MA}.
\bptok{imsref}%
\end{bbook}
%
\endbibitem

\bibitem[\protect\citeauthoryear{Sayrin et al.}{2011}]{Sayetal11}
%
\begin{barticle}[auto:STB|2012/01/27|08:30:54]
\bauthor{\bsnm{Sayrin},~\bfnm{C.}\binits{C.}},
\bauthor{\bsnm{Dotsenko},~\bfnm{I.}\binits{I.}},
\bauthor{\bsnm{Zhou},~\bfnm{X.}\binits{X.}},
\bauthor{\bsnm{Peaudecerf},~\bfnm{B.}\binits{B.}},
\bauthor{\bsnm{Rybarczyk},~\bfnm{T.}\binits{T.}},
\bauthor{\bsnm{Gleyzes},~\bfnm{S.}\binits{S.}},
\bauthor{\bsnm{Rouchon},~\bfnm{P.}\binits{P.}},
\bauthor{\bsnm{Mirrahimi},~\bfnm{M.}\binits{M.}},
\bauthor{\bsnm{Amini},~\bfnm{H.}\binits{H.}},
\bauthor{\bsnm{Brune},~\bfnm{M.}\binits{M.}},
\bauthor{\bsnm{Raimond},~\bfnm{J.~M.}\binits{J.~M.}} \AND
\bauthor{\bsnm{Haroche},~\bfnm{S.}\binits{S.}}
(\byear{2011}).
\btitle{Real-time quantum feedback prepares and stabilizes photon number
states}.
\bjournal{Nature}
\bvolume{477}
\bpages{10376}.
\bptok{imsref}%
\end{barticle}
%
\endbibitem

\bibitem[\protect\citeauthoryear{Schumacher}{1995}]{Sch95}
%
\begin{barticle}[mr]
\bauthor{\bsnm{Schumacher},~\bfnm{Benjamin}\binits{B.}}
(\byear{1995}).
\btitle{Quantum coding}.
\bjournal{Phys. Rev. A (3)}
\bvolume{51}
\bpages{2738--2747}.
\bid{doi={10.1103/PhysRevA.51.2738}, issn={1050-2947}, mr={1328824}}
\bptok{imsref}%
\end{barticle}
%
\endbibitem

\bibitem[\protect\citeauthoryear{Schumacher and
Westmoreland}{1997}]{SchWes97}
%
\begin{barticle}[auto:STB|2012/01/27|08:30:54]
\bauthor{\bsnm{Schumacher},~\bfnm{B.}\binits{B.}} \AND
\bauthor{\bsnm{Westmoreland},~\bfnm{M.~D.}\binits{M.~D.}}
(\byear{1997}).
\btitle{Sending classical information via noisy quantum channels}.
\bjournal{Phys. Rev. A}
\bvolume{56}
\bpages{131138}.
\bptok{imsref}%
\end{barticle}
%
\endbibitem

\bibitem[\protect\citeauthoryear{Shankar}{1994}]{Sha94}
%
\begin{bbook}[mr]
\bauthor{\bsnm{Shankar},~\bfnm{R.}\binits{R.}}
(\byear{1994}).
\btitle{Principles of Quantum Mechanics}, \bedition{2nd} ed.
\bpublisher{Plenum Press}, \baddress{New York}.
\bid{mr={1343488}}
\bptok{imsref}%
\end{bbook}
%
\endbibitem

\bibitem[\protect\citeauthoryear{Shenvi, Kempe and
Whaley}{2003}]{SheKemWha03}
%
\begin{barticle}[auto:STB|2012/01/27|08:30:54]
\bauthor{\bsnm{Shenvi},~\bfnm{N.}\binits{N.}},
\bauthor{\bsnm{Kempe},~\bfnm{J.}\binits{J.}} \AND
\bauthor{\bsnm{Whaley},~\bfnm{K.~B.}\binits{K.~B.}}
(\byear{2003}).
\btitle{Quantum random-walk search algorithm}.
\bjournal{Phys. Rev. A}
\bvolume{67}
\bpages{052307}.
\bptok{imsref}%
\end{barticle}
%
\endbibitem

\bibitem[\protect\citeauthoryear{Shor}{1994}]{Sho94}
%
\begin{bincollection}[mr]
\bauthor{\bsnm{Shor},~\bfnm{Peter~W.}\binits{P.~W.}}
(\byear{1994}).
\btitle{Algorithms for quantum computation: Discrete logarithms and factoring}.
In \bbooktitle{35th {A}nnual {S}ymposium on {F}oundations of {C}omputer
{S}cience ({S}anta {F}e, {NM}, 1994)}
\bpages{124--134}.
\bpublisher{IEEE Comput. Soc. Press}, \baddress{Los Alamitos, CA}.
\bid{doi={10.1109/SFCS.1994.365700}, mr={1489242}}
\bptok{imsref}%
\end{bincollection}
%
\endbibitem

\bibitem[\protect\citeauthoryear{Shor}{1995}]{Sho95}
%
\begin{barticle}[auto:STB|2012/01/27|08:30:54]
\bauthor{\bsnm{Shor},~\bfnm{P.~W.}\binits{P.~W.}}
(\byear{1995}).
\btitle{Scheme for reducing decoherence in quantum computer memory}.
\bjournal{Phys. Rev. A}
\bvolume{52}
\bpages{2493--\break2496}.
\bptok{imsref}%
\end{barticle}
%
\endbibitem

\bibitem[\protect\citeauthoryear{Shor}{1997}]{Sho97}
%
\begin{barticle}[mr]
\bauthor{\bsnm{Shor},~\bfnm{Peter~W.}\binits{P.~W.}}
(\byear{1997}).
\btitle{Polynomial-time algorithms for prime factorization and discrete
logarithms on a quantum computer}.
\bjournal{SIAM J. Comput.}
\bvolume{26}
\bpages{1484--1509}.
\bid{doi={10.1137/S0097539795293172}, issn={0097-5397}, mr={1471990}}
\bptok{imsref}%
\end{barticle}
%
\endbibitem

\bibitem[\protect\citeauthoryear{Steane}{1996a}]{Ste96N1}
%
\begin{barticle}[mr]
\bauthor{\bsnm{Steane},~\bfnm{A.~M.}\binits{A.~M.}}
(\byear{1996}a).
\btitle{Error correcting codes in quantum theory}.
\bjournal{Phys. Rev. Lett.}
\bvolume{77}
\bpages{793--797}.
\bid{doi={10.1103/PhysRevLett.77.793}, issn={0031-9007}, mr={1398854}}
\bptok{imsref}%
\end{barticle}
%
\endbibitem

\bibitem[\protect\citeauthoryear{Steane}{1996b}]{Ste96N2}
%
\begin{barticle}[mr]
\bauthor{\bsnm{Steane},~\bfnm{Andrew}\binits{A.}}
(\byear{1996}b).
\btitle{Multiple-particle interference and quantum error correction}.
\bjournal{Proc. R. Soc. Lond. Ser. A}
\bvolume{452}
\bpages{2551--2577}.
\bid{doi={10.1098/rspa.1996.0136}, issn={0962-8444}, mr={1421749}}
\bptok{imsref}%
\end{barticle}
%
\endbibitem

\bibitem[\protect\citeauthoryear{Szegedy}{2004}]{Sze}
%
\begin{binproceedings}[auto:STB|2012/01/27|08:30:54]
\bauthor{\bsnm{Szegedy},~\bfnm{M.}\binits{M.}}
(\byear{2004}).
\btitle{Quantum speed-up of Markov chain based algorithms}.
In
\bbooktitle{Proc. 45th IEEE Symposium on Foundations of Computer Science}
\bpages{32--41}.
\bpublisher{IEEE Computer Society Press}, \baddress{Los Alamitos, CA}.
\bptok{imsref}%
\end{binproceedings}
%
\endbibitem


\bibitem[\protect\citeauthoryear{Trotter}{1959}]{Tro59}
%
\begin{barticle}[mr]
\bauthor{\bsnm{Trotter},~\bfnm{H.~F.}\binits{H.~F.}}
(\byear{1959}).
\btitle{On the product of semi-groups of operators}.
\bjournal{Proc. Amer. Math. Soc.}
\bvolume{10}
\bpages{545--551}.
\bid{issn={0002-9939}, mr={0108732}}
\bptok{imsref}%
\end{barticle}
%
\endbibitem

\bibitem[\protect\citeauthoryear{Tsirelson}{1980}]{T80}
%
\begin{barticle}[mr]
\bauthor{\bsnm{Tsirelson},~\bfnm{B.~S.}\binits{B.~S.}}
(\byear{1980}).
\btitle{Quantum generalizations of {B}ell's inequality}.
\bjournal{Lett. Math. Phys.}
\bvolume{4}
\bpages{93--100}.
\bid{doi={10.1007/BF00417500}, issn={0377-9017}, mr={0577178}}
\bptok{imsref}%
\end{barticle}
%
\endbibitem

\bibitem[\protect\citeauthoryear{Tulsi}{2008}]{Tul08}
%
\begin{barticle}[auto:STB|2012/01/27|08:30:54]
\bauthor{\bsnm{Tulsi},~\bfnm{A.}\binits{A.}}
(\byear{2008}).
\btitle{Faster quantum-walk algorithm for the two-dimensional spatial search}.
\bjournal{Phys. Rev. A}
\bvolume{78}
\bpages{012310}.
\bptok{imsref}%
\end{barticle}
%
\endbibitem

\bibitem[\protect\citeauthoryear{Vidakovic}{1999}]{Vid99}
%
\begin{bbook}[mr]
\bauthor{\bsnm{Vidakovic},~\bfnm{Brani}\binits{B.}}
(\byear{1999}).
\btitle{Statistical Modeling by Wavelets}.
\bpublisher{Wiley}, \baddress{New York}.
\bid{doi={10.1002/9780470317020}, mr={1681904}}
\bptok{imsref}%
\end{bbook}
%
\endbibitem

\bibitem[\protect\citeauthoryear{von Neumann}{1955}]{von55}
%
\begin{bbook}[mr]
\bauthor{\bparticle{von} \bsnm{Neumann},~\bfnm{John}\binits{J.}}
(\byear{1955}).
\btitle{Mathematical Foundations of Quantum Mechanics}.
\bpublisher{Princeton Univ. Press}, \baddress{Princeton, NJ}.
\bnote{Translated by Robert T. Beyer}.
\bid{mr={0066944}}
\bptok{imsref}%
\end{bbook}
%
\endbibitem


\bibitem[\protect\citeauthoryear{Wang}{1994}]{Wan94}
%
\begin{barticle}[mr]
\bauthor{\bsnm{Wang},~\bfnm{Ya~Zhen}\binits{Y.~Z.}}
(\byear{1994}).
\btitle{Quantum {G}aussian processes}.
\bjournal{Acta Math. Appl. Sin. (Engl. Ser.)}
\bvolume{10}
\bpages{315--327}.
\bid{issn={0168-9673}, mr={1310171}}
\bptok{imsref}%
\end{barticle}
%
\endbibitem

\bibitem[\protect\citeauthoryear{Wang}{2011}]{Wan11}
%
\begin{barticle}[mr]
\bauthor{\bsnm{Wang},~\bfnm{Yazhen}\binits{Y.}}
(\byear{2011}).
\btitle{Quantum {M}onte {C}arlo simulation}.
\bjournal{Ann. Appl. Stat.}
\bvolume{5}
\bpages{669--683}.
\bid{doi={10.1214/10-AOAS406}, issn={1932-6157}, mr={2840170}}
\bptok{imsref}%
\end{barticle}
%
\endbibitem

\bibitem[\protect\citeauthoryear{Wocjan and Abeyesinghe}{2008}]{WocAbe08}
%
\begin{bmisc}[auto:STB|2012/01/27|08:30:54]
\bauthor{\bsnm{Wocjan},~\bfnm{P.}\binits{P.}} \AND
\bauthor{\bsnm{Abeyesinghe},~\bfnm{A.}\binits{A.}}
(\byear{2008}).
\btitle{Speed-up via quantum sampling}.
\bnote{arXiv:\arxivurl{0804.4259v3}[quant-ph]}.
\bptok{imsref}%
\end{bmisc}
%
\endbibitem


\bibitem[\protect\citeauthoryear{Wootters and Zurek}{1982}]{WooZur82}
%
\begin{barticle}[auto:STB|2012/01/27|08:30:54]
\bauthor{\bsnm{Wootters},~\bfnm{W.~K.}\binits{W.~K.}} \AND
\bauthor{\bsnm{Zurek},~\bfnm{W.~H.}\binits{W.~H.}}
(\byear{1982}).
\btitle{A single quantum cannot be cloned}.
\bjournal{Nature}
\bvolume{299}
\bpages{802--803}.
\bptok{imsref}%
\end{barticle}
%
\endbibitem

\bibitem[\protect\citeauthoryear{Yu}{2001}]{autokey82}
%
\begin{binproceedings}[mr]
\bauthor{\bsnm{Yu},~\bfnm{A.~B.}\binits{A.~B.}}
(\byear{2001}).
\btitle{Campbell-Hausdorff formula}.
In \bbooktitle{Encyclopaedia of Mathematics}
(\beditor{\bfnm{M.}\binits{M.}~\bsnm{Hazewinkel}}, ed.).
\bpublisher{Kluwer Academic}, \baddress{Dordrecht}.
\bptok{imsref}%
\end{binproceedings}
%
\endbibitem

\bibitem[\protect\citeauthoryear{Zalka}{1998}]{Zal98}
%
\begin{barticle}[auto:STB|2012/01/27|08:30:54]
\bauthor{\bsnm{Zalka},~\bfnm{C.}\binits{C.}}
(\byear{1998}).
\btitle{Simulating a quantum systems on a quantum computer}.
\bjournal{R. Soc. Lond. Philos. Trans. Ser. A Math. Phys. Eng. Sci.}
\bvolume{454}
\bpages{313--322}.
\bptok{imsref}%
\end{barticle}
%
\endbibitem

\end{thebibliography}
\end{document}